\documentclass[prd,preprint,tightenlines,floatfix,showpacs,preprintnumbers,nofootinbib,eqsecnum,cernpreprint]{revtex4-1}

\pdfoutput=1

\usepackage[dvips,final]{graphicx}
\usepackage{amssymb}
\usepackage{amsmath}
\usepackage{amsfonts}
\usepackage{epsfig}
\usepackage{bm}
\usepackage{comment}
\usepackage{float}
\usepackage[utf8]{inputenc}
\usepackage{caption}
\usepackage{subcaption}
\usepackage{hhline}
\usepackage[normalem]{ulem}

\usepackage[dvipsnames,table,xcdraw]{xcolor}
\usepackage{array}
\usepackage{multirow}   
\usepackage{booktabs}   
\usepackage{bbold}
\usepackage{hyperref}
\hypersetup{colorlinks=true, linkcolor=teal, urlcolor=blue, citecolor=red}
\usepackage{enumerate}
\usepackage{mathtools}

\usepackage{physics} 
\usepackage{makecell} 

\usepackage{cancel}

\usepackage{feynmp-auto} 

\usepackage[capitalise]{cleveref}

\crefname{section}{Sec.}{Secs.}
\crefname{table}{Tab.}{Tabs.}
\crefname{figure}{Fig.}{Figs.}
\crefname{equation}{Eq.}{Eqs.}
\crefname{appendix}{Appendix\ }{Appendix\ }

\definecolor{bostonuniversityred}{rgb}{0.8, 0.0, 0.0}

\newcommand{\redBU}[0]{\color{bostonuniversityred}}

\newcommand\varpm{\mathbin{\vcenter{\hbox{%
  \oalign{\hfil$\scriptstyle\hspace{-0.1ex}+\hspace{-0.1ex}$\hfil\cr
          \noalign{\kern-.5ex}
          $\scriptscriptstyle({-})$\cr}%
}}}}

\DeclareSymbolFont{myletters}{OML}{ztmcm}{m}{it}
\DeclareMathSymbol{\uplambda}{\mathord}{myletters}{"15}

\begin{document}

\begin{flushright}
CERN-TH-2022-194\\
\vskip1cm
\end{flushright}

\title{Collider phenomenology of new neutral scalars\\ 
in a flavoured multi-Higgs model}

\author{P.M. Ferreira$^{1,2}$}
\email{pedro.ferreira@fc.ul.pt}

\author{João Gonçalves$^{3}$}
\email{jpedropino@ua.pt}

\author{Antonio P. Morais$^{3,4}$}
\email{a.morais@cern.ch}

\author{Ant\'onio Onofre$^{5}$}
\email{onofre@fisica.uminho.pt}

\author{Roman Pasechnik$^{6}$}
\email{Roman.Pasechnik@thep.lu.se}

\author{Vasileios Vatellis$^{3}$}
\email{vasileios.vatellis@gmail.com}

\affiliation{
{$^1$\sl
Instituto~Superior~de~Engenharia~de~Lisboa~---~ISEL, 1959-007~Lisboa, Portugal
}\\
{$^2$\sl
Centro~de~F\'{\i}sica~Te\'orica~e~Computacional, Faculdade~de~Ci\^encias,
Universidade~de~Lisboa, Campo Grande, 1749-016~Lisboa, Portugal
}\\
{$^3$\sl 
Departamento de F\'isica, Universidade de Aveiro and CIDMA, Campus de Santiago, 
3810-183 Aveiro, Portugal
}\\
{$^4$\sl 
Theoretical Physics Department, CERN, 1211 Geneva 23, Switzerland
}\\
{$^5$\sl
Departamento de F\'{i}sica da Universidade do Minho, 4710-057 Braga, Portugal
}\\
{$^6$\sl
Department of Astronomy and Theoretical Physics, Lund
University, SE-223 62 Lund, Sweden
}}

\begin{abstract}
\vspace{0.5cm}
In this work, we propose and explore for the first time a new collider signature of heavy neutral scalars typically found in many distinct classes of multi-Higgs models. This signature, particular relevant in the context of the Large Hadron Collider (LHC) measurements, is based on a topology with two charged leptons and four jets arising from first and second generation quarks. As an important benchmark scenario of the multi-Higgs models, we focus on a recently proposed Branco-Grimus-Lavoura (BGL) type model enhanced with an abelian $\mathrm{U(1)}$ flavour symmetry and featuring an additional sector of right-handed neutrinos. We discuss how kinematics of the scalar fields in this model can be used to efficiently separate the signal from the dominant backgrounds and explore the discovery potential of the new heavy scalars in the forthcoming LHC runs. The proposed method can be applied for analysis of statistical significance of heavy scalars' production at the LHC and future colliders in any multi-Higgs model.
\end{abstract}


\maketitle
\tableofcontents

\newpage
\maketitle

\section{Introduction}\label{sec:Intro}

The current basis for our understanding of particle physics is leaning on the theoretical framework of the Standard Model (SM), which was ultimately confirmed by the discovery of the Higgs boson \cite{Chatrchyan:2012xdj,ATLAS:2012yve}, whose properties closely match the SM expectations. Despite this, an explanation for neutrino masses, for the observed hierarchies in the fermionic sector and for the existence of dark matter cannot be accommodated in the SM. Possible solutions to the aforementioned questions require a New Physics (NP) framework that typically incorporates new scalar fields, including both electrically charged and neutral Higgs states with distinct CP properties.

The presence of multiple beyond the SM (BSM) particles can lead to interesting phenomenology at collider experiments, with a multitude of possible different final states. Indeed, over the years, various searches have been conducted by the experimental community, including decays that involve vector bosons \cite{ATLAS:2021uiz,ATLAS:2020fry,ATLAS:2020tlo}, multi-jets \cite{ATLAS:2020jgy,CMS:2020osd} and charged leptons \cite{ATLAS:2020zms,ATLAS:2021ldb,CMS:2020ffa}. However, the presence of new scalars that interact and/or mix with one another lead to additional topologies which have not been covered by experiments. In particular, decay chains featuring various BSM fields in internal propagators offer a solid physics case to test generic multi-Higgs models at the LHC. In this regard, while there have been a number of recent searches \cite{ATLAS:2020pcy,ATLAS:2020gxx}, they have been limited in scope, especially when compared to topologies where a BSM field decays immediately into a pair of SM particles. Therefore, the proposal in this article aims at filling such a gap and enlarging the explored parameter space in view of the LHC run-III, which is already collecting data, and of the upcoming High-Luminosity (HL) upgrade.

In this regard, we consider a particular example of a multi-Higgs model featuring the Branco-Grimus-Lavoura (BGL) flavour structure \cite{Branco:1996bq} as a benchmark for our phenomenological analysis. Besides the two Higgs doublets and a complex singlet scalar charged under a global flavour $\mathrm{U(1)}$ symmetry, the model also incorporates an additional sector with three generations of right-handed neutrinos and a type-I seesaw mechanism for generation of light neutrino masses which was recently explored by some of the authors in \cite{Ferreira:2022zil}. In this work, a comprehensive phenomenological analysis of this model has been performed, where electroweak (EW) precision, Higgs, flavour and the existing collider observables were scrutinized. A selection of the phenomenologically validated points are then used as benchmark scenarios for exploring the discovery potential of the heavy neutral BSM scalars in the considered NP framework.

This article is organised as follows. In Sec.~\ref{sec:Model} we briefly discuss a particular benchmark NP model that is chosen for our numerical analysis. In Sec.~\ref{section:LHC_pheno} we give a general overview of the current experimental status regarding the search for heavy BSM scalar states at collider experiments. In Sec.~\ref{sec:BSM_scalars}, we discuss a number of processes involving such states that have been so far a subject of little or no attention in the literature. Here, we propose a new promising signature featuring a final state with four jets and a pair of charged leptons and with multiple neutral BSM scalars in internal propagators. In Sec.~\ref{sec:Collider}, we present the details of the newly developed methodology while in Sec.~\ref{sec:numerical_results} we calculate the statistical significance for discovery/exclusion of new Higgs states. Finally, in Sec.~\ref{sec:Conclusion}, we conclude and summarise our results.

\section{Benchmark model}\label{sec:Model}

The recently proposed BGL-like Next-to-Minimal Two Higgs Doublet Model (BGL-NTHDM) has been thoroughly discussed both from the theoretical and phenomenological points of view in \cite{Ferreira:2022zil}. Here, for completeness of information, we briefly present its key features. The BGL structure results from a global flavour $\mathrm{U(1)}$ symmetry, broken once the Higgs doublets and the scalar singlet develop vacuum expectation values (VEVs) \cite{Branco:1996bq}. The allowed Yukawa interactions read as 
\begin{equation}\label{eq:Yukawa_sector}
\begin{aligned}
-\mathcal{L}_\mathrm{Yukawa}&=\overline{q_L^0}\Gamma_a\Phi^a d_R^0+\overline{q_L^0}\Delta_a\tilde{\Phi}^a u_R^0 +\overline{\ell_L^0}\Pi_a\Phi^a e_R^0+\overline{\ell_L^0}\Sigma_a\tilde{\Phi}^a \nu_R\\
&+\frac{1}{2}\overline{\nu_R^c}\left(\mathrm{A}+\mathrm{B}S+\mathrm{C}S^\ast\right)\nu_R+\mathrm{H.c.}\,,
\end{aligned}
\end{equation}
where the index $a=1,2$ runs over the two doublets $\Phi^a$ and all the matrices are written in the flavour basis, with $\Gamma_a$, $\Delta_a$ being the Yukawa matrices for down-quarks, up-quarks and $\Pi_a$, $\Sigma_a$ are the charged lepton and neutrino Yukawa matrices, respectively. Besides, here $B$ and $C$ are the Majorana-like Yukawa couplings with a complex SU(2) singlet scalar field $S$, whereas $A$ is a Majorana mass term for the right-handed neutrinos.  We adopt the notation where $\tilde{\Phi}\equiv i \sigma_2{\Phi}^*$ and the superscript $0$ indicates that the fields are written in the gauge basis. The textures of the quark-sector Yukawa matrices are given as
\begin{equation}\label{eq:textures_Yukawas_quarks}
\begin{aligned}
\Gamma_1&:
\begin{pmatrix}
\times& \times & \times
\\
\times &\times & \times
\\
 \makebox[\widthof{$\times$}][c]{0} & \makebox[\widthof{$\times$}][c]{0}&  \makebox[\widthof{$\times$}][c]{0}
\end{pmatrix} \,,
\hspace{2mm}
\Gamma_2:
\begin{pmatrix}
 \makebox[\widthof{$\times$}][c]{0}&  \makebox[\widthof{$\times$}][c]{0}& \makebox[\widthof{$\times$}][c]{0}
\\
 \makebox[\widthof{$\times$}][c]{0}&    \makebox[\widthof{$\times$}][c]{0}& \makebox[\widthof{$\times$}][c]{0}
\\
\times & \times & \times
\end{pmatrix} \,,
\hspace{2mm}
\Delta_1:
\begin{pmatrix}
\times& \times &\makebox[\widthof{$\times$}][c]{0}
\\
\times&\times & \makebox[\widthof{$\times$}][c]{0}
\\
\makebox[\widthof{$\times$}][c]{0} &\makebox[\widthof{$\times$}][c]{0} &\makebox[\widthof{$\times$}][c]{0}
\end{pmatrix} \,,
\hspace{1,5mm}
\Delta_2:
\begin{pmatrix}
\makebox[\widthof{$\times$}][c]{0} &\makebox[\widthof{$\times$}][c]{0} & \makebox[\widthof{$\times$}][c]{0} 
\\
\makebox[\widthof{$\times$}][c]{0}&\makebox[\widthof{$\times$}][c]{0}  & \makebox[\widthof{$\times$}][c]{0}
\\
\makebox[\widthof{$\times$}][c]{0} & \makebox[\widthof{$\times$}][c]{0}& \times
\end{pmatrix} \,,
\end{aligned}
\end{equation}
while those of the charged and neutral leptons read as
\begin{align}\label{eq:textures_Yukawas_leptons}
\begin{split}
\Pi_1,\Sigma_1,B=
\begin{pmatrix}
\times&\times& \makebox[\widthof{$\times$}][c]{0}\\
\times&\times& \makebox[\widthof{$\times$}][c]{0}\\
 \makebox[\widthof{$\times$}][c]{0}& \makebox[\widthof{$\times$}][c]{0}& \makebox[\widthof{$\times$}][c]{0}
\end{pmatrix}\,,\quad
\Pi_2,\Sigma_2=
\begin{pmatrix}
 \makebox[\widthof{$\times$}][c]{0}& \makebox[\widthof{$\times$}][c]{0}& \makebox[\widthof{$\times$}][c]{0}\\
 \makebox[\widthof{$\times$}][c]{0}& \makebox[\widthof{$\times$}][c]{0}& \makebox[\widthof{$\times$}][c]{0}\\
 \makebox[\widthof{$\times$}][c]{0}& \makebox[\widthof{$\times$}][c]{0}&\times
\end{pmatrix}\,\quad
A=\mathbb{0}\,,\quad
C=
\begin{pmatrix}
 \makebox[\widthof{$\times$}][c]{0}& \makebox[\widthof{$\times$}][c]{0}&\times\\
 \makebox[\widthof{$\times$}][c]{0}& \makebox[\widthof{$\times$}][c]{0}&\times\\
\times&\times& \makebox[\widthof{$\times$}][c]{0}
\end{pmatrix}\,.
\end{split}
\end{align}
In the gauge basis, the fermion mass matrices can be cast as
\begin{equation}\label{eq:tree_level_masses}
\begin{aligned}
&M_u^0\equiv\frac{1}{\sqrt{2}}\left(v_1 \Delta_1+v_2\Delta_2\right), \\
&M_d^0=\frac{1}{\sqrt{2}}\left(v_1 \Gamma_1+v_2\Gamma_2\right), \\
&M_e^0=\frac{1}{\sqrt{2}}\left(v_1 \Pi_1+v_2\Pi_2\right),
\end{aligned}
\end{equation}
where $u,d,e$ denote up, down quarks and charged leptons, while $v_1$ and $v_2$ are the VEVs arising from the first and second Higgs doublets, respectively. Such mass matrices can be rotated to the physical basis via bi-unitary transformations. 

Generically, each fermion $f=u,d,e$ can be diagonalized as follows,
\begin{equation}\label{eq:diagonal_fermions}
N_f = U_{f\mathrm{L}}^\dagger M_f^0 \, U_{f\mathrm{R}} \,,
\end{equation}
where $U$ are the unitary matrices, with the subscripts L(R) denoting left(right) chirality, and $N_f$ are the diagonal fermion mass matrices. It follows from the textures in Eqs.~\eqref{eq:textures_Yukawas_quarks} and \eqref{eq:textures_Yukawas_leptons} that both the charged lepton and up-quark Yukawa matrices can be simultaneously diagonalized, and therefore there are no tree-level FCNCs for both sectors. For down-quarks, $\Gamma_1$ and $\Gamma_2$ can not be simultaneously diagonalized and therefore FCNCs will be present readily at tree-level, mediated by new BSM scalars. As in the standard BGL construction \cite{Branco:1996bq}, the effect of the global flavour symmetry results in a suppression of FCNCs by the off-diagonal elements of the Cabibbo-Kobayashi-Maskawa (CKM) matrix. 

While not particularly relevant for the current study, the model also features an additional sector of massive right-handed neutrinos, which, for completeness, is briefly outlined below (for more details, see \cite{Ferreira:2022zil}). Defining the neutrino fields as
\begin{equation}\label{eq:neutrino_fields}
n_L^0\equiv
\begin{pmatrix}
\nu_L^0\\
\nu_R^c
\end{pmatrix} \,,
\end{equation}
the mass matrix can be cast in the standard seesaw format as
\begin{equation}\label{eq:neutrino_mass_matrix}
\mathcal{M}\equiv
\begin{pmatrix}
\mathbb{0}&m_D\\
m_D^T&M_R
\end{pmatrix} \,,
\end{equation}
where one defines 
\begin{equation}\label{eq:matrix_elements_neutrinos}
\begin{aligned}
m_D&\equiv\frac{1}{\sqrt{2}}\left(v_1\Sigma_1+v_2\Sigma_2\right) \,,\\
M_R&\equiv\mathrm{A}+\frac{v_S}{\sqrt{2}}\left(\mathrm{B}+\mathrm{C}\right) \,,
\end{aligned}
\end{equation}
and $v_S$ is the VEV in the real component of the complex scalar singlet. As it was noted in our previous work \cite{Ferreira:2022zil}, the model possesses enough freedom to simultaneously fit the Pontecorvo-Maki-Nakagawa-Sakata (PMNS) neutrino mixing matrix and the active neutrino mass differences.

The scalar potential is defined as $V=V_0 + V_1$, with
\begin{equation}\label{eq:Scal_Pote_1}
\begin{aligned}
V_0 =\quad  & \mu_{a}^2|\Phi^{a}|^2 + \lambda_{a}|\Phi^{a}|^4 + \lambda_{3}|\Phi_{1}|^2|\Phi_{2}|^2 + \lambda_{4}|\Phi_{1}^{\dagger}\Phi_{2}|^2  + {\mu_S}^{2}|S|^2 +  \lambda_{1}^\prime|S|^4 + \lambda_{2}^\prime|\Phi_{1}|^2|S|^2\\  & + \lambda_{3}^\prime|\Phi_{2}|^2|S|^2 \,,\\
V_1 =\quad & \mu_{3}^{2} \Phi_{2}^{\dagger}\Phi_1  + \frac{1}{2}\mu_{b}^{2} S^{2} + a_{1} \Phi_{1}^{\dagger}\Phi_2 S + a_{2} \Phi_{1}^{\dagger}\Phi_2 S^{\dagger} +  a_{3} \Phi_{1}^{\dagger}\Phi_2 S^{2} + \mathrm{H.c.}\, .
\end{aligned}\end{equation}
While $V_0$ contains the quartic couplings ($\lambda_{1,2,3,4},\lambda^{\prime}_{1,2,3}$) and the quadratic mass terms ($\mu_{1,2,S}^2$), $V_1$ contains the soft $\mathrm{U(1)}$-breaking interactions ($\mu_3,\mu_b,a_{1,2}$), as well as the quartic coupling between the singlet and the two Higgs doublets ($a_3$) invariant under the $\mathrm{U(1)}$ flavour symmetry. Once the scalars develop VEVs, six physical states emerge including three CP-even neutral scalars $H_1$, $H_2$ and $H_3$, with $H_1$ corresponding to the SM-like Higgs boson, two CP-odd neutral scalars $A_2$ and $A_3$, and a charged scalar $H^\pm$.

\section{Heavy Higgs partners at the LHC}\label{section:LHC_pheno}

Multi-Higgs models provide a plethora of new scalar states that can be either charged or neutral and possess CP-even or CP-odd properties, leading to characteristic signatures in collider measurements. A summary of the most recent searches, performed in years of 2020 and 2021 at both the CMS and ATLAS experiments, is shown in Tab.~\ref{tab:CMS_ATLAS}. Various combinations of final states have been searched for, particularly, for the neutral fields commonly referred to as Higgs partners, $H$ and $A$, including final states with light jets, $b$-jets and charged leptons. It is also interesting to note that searches involving decays into other BSM fields are also included, such as $A\rightarrow H\mathrm{Z^0}$ in \cite{ATLAS:2020gxx} and $H\rightarrow AA$ in \cite{CMS:2020ffa}. These channels are of particular relevance for the Higgs partners' search since interactions between different BSM scalars are common to most (if not all) multi-Higgs extensions of the SM. We do note that in \cite{CMS:2020ffa} the search focused in the low-mass regime for the CP-odd scalar, complementing the high-mass regime in ATLAS for the processes $A\rightarrow \tau^+\tau^-$ and $A\rightarrow \tau^+\tau^-b\bar{b}$ \cite{ATLAS:2020zms} and $A\rightarrow \gamma\gamma$ \cite{ATLAS:2021uiz}, has been performed. Note, the CP-odd ($A$) and CP-even ($H$) scalars often share the same final states as highlighted in Tab.~\ref{tab:CMS_ATLAS}.
\begin{table}[htb!] \centering
\captionsetup{justification=raggedright,singlelinecheck=false}
\caption{A summary of the most recent searches conducted between the years of 2020 and 2021 by the ATLAS and CMS experiments at the LHC. Here, $H_1$ is the SM-like Higgs boson, $\bm{A}$ represents a neutral pseudoscalar, $\bm{H}$ is a CP-even neutral scalar and $\bm{H^\pm}$ is a charged scalar. Here, ``BR'' stands for the corresponding branching ratio. As for the production mechanisms of the Higgs states, ``ggF'' indicates the gluon-gluon fusion, while ``VBF'' corresponds to the vector boson fusion.}
\label{tab:CMS_ATLAS}
\begin{small}
\begin{tabular}{@{}ccccc@{}}\toprule
\textbf{Scalar field} & \textbf{Decay channel} & \textbf{Mass limits (GeV)} & \textbf{Comments} & \textbf{Refs.} \\ \hline
\multirow{13}{*}{\textbf{$\bm{A}$}} & $A\rightarrow \tau^+\tau^-$ & $[200, 2500]$ & \makecell{Limits given in \\ terms of $\sigma\times \mathrm{BR}$} & \cite{ATLAS:2020zms} \\[0.5em] \cline{2-5}
 &  $A \rightarrow \tau^+\tau^-b\bar{b}$ & $[200, 2500]$ & \makecell{Limits given in \\ terms of $\sigma\times \mathrm{BR}$} & \cite{ATLAS:2020zms} \\[0.5em] \cline{2-5} 
 & $H_1\rightarrow A\mathrm{Z^0}$ & $[0.5, 4.0]$ & \makecell{Hadronic decays with \\ $\mathrm{BR}(A\rightarrow gg)=1$ or \\ $\mathrm{BR}(A\rightarrow s\bar{s})=1$} & \cite{ATLAS:2020pcy} \\[0.5em] \cline{2-5}
 & $AA \rightarrow b\bar{b}b\bar{b}$ & $[20, 60]$ & \makecell{Limits given in \\ terms of $\sigma\times \mathrm{BR}$ \\ Associated $\mathrm{Z^0}$ production} & \cite{ATLAS:2020ahi} \\[0.5em] \cline{2-5}
 & $A\rightarrow H\mathrm{Z^0}$& $-$ & \makecell{Limits $m_H$ vs. $m_A$ \\ Multiple channels \\ $2\ell2b$, $2\ell4j$, $2\ell4b$} & \cite{ATLAS:2020gxx} \\[0.5em] \cline{2-5}
 & $A\rightarrow \gamma\gamma$& $[160, 2800]$ & \makecell{Limits given in \\ terms of $\sigma\times \mathrm{BR}$} & \cite{ATLAS:2021uiz} \\[0.5em] \hline \hline
\multirow{14}{*}{\textbf{$\bm{H}$}} & $H\rightarrow \tau^+\tau^-$ & $[200, 2500]$ & \makecell{Limits given in \\ terms of $\sigma\times \mathrm{BR}$} & \cite{ATLAS:2020zms} \\[0.5em]  \cline{2-5}
 &  $H \rightarrow \tau^+\tau^-b\bar{b}$ & $[200, 2500]$ & \makecell{Limits given in \\ terms of $\sigma\times \mathrm{BR}$} & \cite{ATLAS:2020zms} \\[0.5em]  \cline{2-5}
 & $HH \rightarrow b\bar{b}b\bar{b}$& $[260, 1000]$ & \makecell{Vector-boson fusion \\ Coupling constraints} & \cite{ATLAS:2020jgy} \\[0.5em]  \cline{2-5}
 & $H \rightarrow V V$& \makecell{$[300, 3200]~\mathrm{ggF}$ \\ $[300, 760]~\mathrm{VBF}$ \\ $[300, 2000]~\mathrm{ggF}$ } & \makecell{First two for Kaluza-Klein (KK)\\ massive gravitons, third for radion. \\ $V$ indicates vector boson} & \cite{ATLAS:2020fry} \\[0.5em]  \cline{2-5}
  & $H \rightarrow \mathrm{Z^0}{Z^0}$& $[400, 2000]$ & \makecell{Various widths assumptions \\ VBF and gluon fusion \\ Fully and semi-leptonic} & \cite{ATLAS:2020tlo} \\[0.5em]  \cline{2-5}
 & $H\rightarrow \gamma\gamma$& $[160, 2800]$ & \makecell{Limits given in \\ terms of $\sigma\times \mathrm{BR}$} & \cite{ATLAS:2021uiz} \\[0.5em] \cline{2-5}
 & $H(H_1)\rightarrow AA$& \makecell{$[16, 62]$ \\ $[15,60]$ \\ $[3.6,21]$} & \makecell{$H_1\rightarrow AA\rightarrow b\bar{b}\mu^+\mu^-$ \\ $H_1\rightarrow AA\rightarrow \ell^+\ell^-\ell^+\ell^-$ \\ $H(H_1)\rightarrow AA\rightarrow\mu^+\mu^-\tau^+\tau^-$} & \makecell{\cite{ATLAS:2021hbr} \\ \cite{ATLAS:2021ldb} \\ \cite{CMS:2020ffa}} \\[0.5em] \hline \hline
\multirow{5}{*}{\textbf{$\bm{H^\pm}$}} & $p p \rightarrow tbH^+$ & \makecell{$[200, 2000]$ \\ $[200, 3000]$} & \makecell{In both: $H^+ \rightarrow tb$ \\ Constraints of $m_H^\pm$ vs. $\tan \beta$ (both) \\ Limits as $\sigma \times \mathrm{BR}$ (both)} & \makecell{\cite{ATLAS:2021upq} \\ \cite{CMS:2020imj}} \\[0.5em] \cline{2-5}
 & $H^\pm \rightarrow W^\pm  \mathrm{Z^0}$& $[200, 1500]$ & \makecell{Considers VBF production \\ Limits as $\sigma \times \mathrm{BR}$ } &  \cite{CMS:2021wlt} \\[0.5em] \cline{2-5}
 & $H^\pm \rightarrow cs$& $[80, 160]$ & \makecell{Assumes $\mathrm{BR}(H^\pm \rightarrow cs) = 1$ \\ Limits as $\mathrm{BR}(t \rightarrow H^+ b)$ vs $m_{H^+}$ } &  \cite{CMS:2020osd} \\[0.5em] \hline \hline
\end{tabular}
\end{small}
\end{table}

Regarding the charged Higgs bosons, some searches have also been reported recently, with a focus on the $tbH^+$ vertex, either through decay into $tb$ \cite{ATLAS:2021upq, CMS:2020imj} or considering a charged Higgs state produced via a top/anti-bottom pair \cite{CMS:2020osd}. An additional search focusing on the vector-boson fusion (VBF) mechanism was also performed by CMS \cite{CMS:2021wlt}. The top/anti-bottom channel appears to be the preferred channel in the search for charged Higgs bosons, corroborated by previous searches in \cite{ATLAS:2018ntn,ATLAS:2015nkq,ATLAS:2016avi,CMS:2018dzl,CMS:2015yvc,CMS:2015yvc,ATLAS:2018gfm,CMS:2019bfg}. Indeed, these studies indicate that for a charged Higgs state with mass {\color{red} below} that of the top quark, it will be predominantly produced via top quark decays, whereas heavy charged scalars would be typically produced in association with a top quark \cite{LHCHiggsCrossSectionWorkingGroup:2016ypw,Branco:2011iw}. For masses {\color{red} below} the kinematic threshold for the production of a top quark, the decay modes $H^\pm \rightarrow \tau^\pm \nu_\tau$ become dominant \cite{LHCHiggsCrossSectionWorkingGroup:2016ypw,Branco:2011iw}.

Besides the searches in \cite{ATLAS:2021uiz,CMS:2020ffa} (see also Tab.~\ref{tab:CMS_ATLAS}), the preference in the literature has been given to the BSM scenarios where new scalars purely decay into SM states. However, multi-Higgs models enable interactions among different BSM scalars such that a richer set of final states involving multiple scalars is possible and must be considered as discussed below.

\section{LHC phenomenology of BSM scalars}\label{sec:BSM_scalars}

The model considered in this article was previously validated in \cite{Ferreira:2022zil}, where several points consistent with EW, Higgs, collider and flavour physics observables were found\footnote{Some of the data and the UFO model (both \texttt{python2} and \texttt{python3} versions) are publicly available in one of the author's GitHub page (see \url{https://github.com/Mrazi09/BGL-ML-project}). The numerical values for the various couplings/masses is also shown in appendix~\ref{app:Numerical_benchmarks}.}. In particular, it was shown that scenarios with light scalars, i.e. being not too far above the Higgs boson mass, are still allowed and can be potentially probed at the LHC run-III or its high-luminosity (HL) phase. Testing these possibilities is, therefore, of phenomenological interest. Our focus here is on either single or double-production channels for the New Physics states present in the model, in particular, charged-, ($H^\pm$), neutral- ($H_2$ and $H_3$) and pseudo-scalars ($A_2$ and $A_3$). While a multitude of topologies for collider searches of these states can be proposed, our current analysis is dedicated to a single production process featuring the final states of four jets and two charged leptons and leaving other possible topologies for a future work.

\subsection{CP-even neutral scalars}\label{subsec:CPevent}

The model considered in this work features two CP-even scalars $H_{2,3}$ with different couplings, masses (satisfying $m_{H_3}>m_{H_2}$) and branching ratios (BRs) for separate decay modes. In what follows, both $H_{2,3}$ are assumed to be heavier than the Higgs boson (denoted as $H_1$). Here, we are only interested in channels with sizeable BRs by requiring them to be greater than $20\%$. As it was previously shown in \cite{Ferreira:2022zil}, the decay channels with small BRs typically result in small cross sections, even below the allowed sensitivity of the HL phase of the LHC. With this in mind, we have selected six preferred scenarios among the valid parameter-space points of \cite{Ferreira:2022zil}, for which the characteristics of the next-to-lightest Higgs state $H_2$ -- the mass and the dominant BR -- read as
\begin{equation}\label{eq:BRS}\nonumber
\begin{aligned}
&\text{\underline{Benchmark H1:}}\quad M_{H_2} = 599.09~\mathrm{GeV}, \quad \mathrm{BR}(H_2 \rightarrow A_2 \mathrm{Z^0}) = 90.7\%\, , \\
&\text{\underline{Benchmark H2:}}\quad M_{H_2} = 286.92~\mathrm{GeV}, \quad \mathrm{BR}(H_2 \rightarrow H_1 H_1) = 66.7 \%\, , \\  
&\text{\underline{Benchmark H3:}}\quad M_{H_2} = 527.55~\mathrm{GeV}, \quad \mathrm{BR}(H_2 \rightarrow A_2 \mathrm{Z^0}) = 52.1\%\, , \\
&\text{\underline{Benchmark H4:}}\quad M_{H_2} = 397.45~\mathrm{GeV}, \quad \mathrm{BR}(H_2 \rightarrow A_2 \mathrm{Z^0}) = 41.7\%\, , \\
&\text{\underline{Benchmark H5:}}\quad M_{H_2} = 215.56~\mathrm{GeV}, \quad \mathrm{BR}(H_2 \rightarrow c \bar{c}) = 56.0\%\, , \\
&\text{\underline{Benchmark H6:}}\quad M_{H_2} = 402.81~\mathrm{GeV}, \quad \mathrm{BR}(H_2 \rightarrow A_2 \mathrm{Z^0}) = 89.6\%\, ,
\end{aligned}
\end{equation}
that are worth of a dedicated phenomenological analysis. Note, however, that single $H_2$ production processes with the dominant decay channels corresponding to those of the benchmarks H1, H2, H3, H4 and H6 have already been studied at the LHC. In particular, the most recent $H_2$ searches in the H1, H3, H4 and H6 channels are already indicated in Tab.~\ref{tab:CMS_ATLAS}. Besides, final states with at least two b-jets were considered for the benchmark H2 \cite{ATLAS:2018uni,ATLAS:2018hqk,ATLAS:2018dpp,CMS:2018vjd,CMS:2018tla,CMS:2017rpp}. Notice that the benchmark H5, whose BR into a pair of charm quarks is $56\%$, represents a rather attractive physics case motivating the searches involving light jets in the final state. Of course, a fully hadronic signal is not optimal in the context of the LHC since the SM multi-jet background is expected to dominate over the signal. While fully hadronic final states have been searched for at the ATLAS experiment (see e.g.~\cite{ATLAS:2022ozf,ATLAS:2020jgy,ATLAS:2018rnh,ATLAS:2019npw}), such a signature can become a lot cleaner \textit{e.g.}~at future lepton colliders.

\begin{figure}[htb!]
    \captionsetup{justification=raggedright,singlelinecheck=false}
    \centering
    \includegraphics[width=0.39\textwidth]{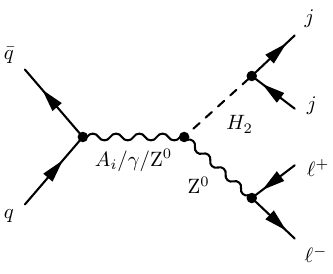}
    \includegraphics[width=0.46\textwidth]{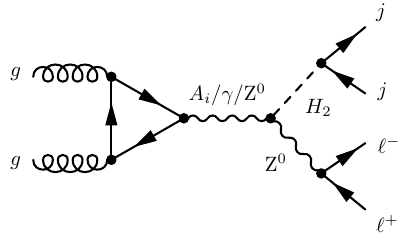} \\
    \includegraphics[width=0.46\textwidth]{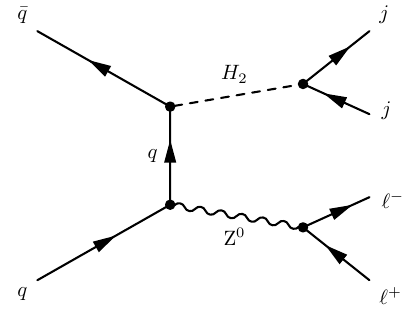}
\caption{The LO Feynman diagrams for production of the single $H_2$ CP-even scalar in the $s$-channel (top two diagrams) and the $t$-channel (bottom diagram) processes. Here, $q$ denotes the SM quarks, $g$ corresponds to a gluon, and $A_{i=2,3}$ -- to CP-odd scalars, while $j$ stands for a jet originating from the physical quarks of first and second generations $u,s,d,c$ and their anti-particles.}
    \label{fig:H2_singleproduction}
\end{figure}

\begin{figure}[htb!]
    \captionsetup{justification=raggedright,singlelinecheck=false}
    \centering
    \includegraphics[width=0.39\textwidth]{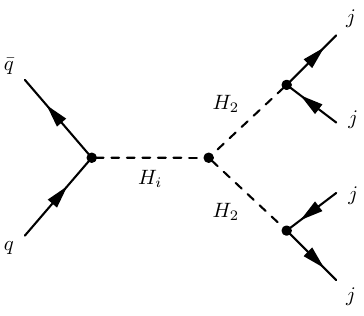}
    \includegraphics[width=0.46\textwidth]{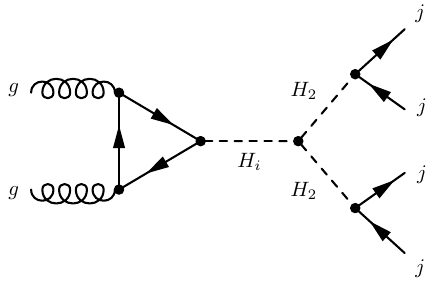} \\
    \includegraphics[width=0.46\textwidth]{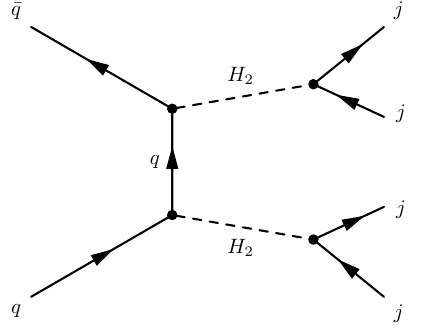}
    \caption{The LO Feynman diagrams for pair-production of the $H_2$ CP-even scalars in the $s$-channel (top two diagrams) and the $t$-channel (bottom diagram) processes. Here, $q$ denotes the SM quarks, $g$ corresponds to a gluon, and $H_{i=1,2,3}$ -- to CP-even scalars, while $j$ stands for a jet originating from the physical quarks of first and second generations $u,s,d,c$ and their anti-particles.}
    \label{fig:H2_pairproduction}
\end{figure}

With this in mind, one may consider two possible cases -- double and single $H_2$ production -- separately. For the latter case, the focus is typically on associated production with a $\mathrm{Z^0}$ boson (see Fig.~\ref{fig:H2_singleproduction}) subsequently decaying into a pair of leptons, thus minimizing the relevant QCD backgrounds. For the former case, all leading-order (LO) diagrams are shown in Fig.~\ref{fig:H2_pairproduction}, where both $s$ and $t$-channels contribute to the matrix element. Additionally, the gluon-gluon fusion mode is relevant and needs also to be included in the calculation. The major ingredients of the background are represented by di-boson, $\mathrm{V}+\mathrm{jets}$ and $t\bar{t}$ production modes that will be carefully considered in what follows.

We can now focus our attention on the heaviest scalar state, $H_3$. Again, we are interested in points with BRs higher than 20\%. Considering the same benchmark scenarios as above
\begin{equation}\label{eq:BRS_1}\nonumber
\begin{aligned}
&\text{\underline{Benchmark H1:}}\quad M_{H_3} = 907.61~\mathrm{GeV}, \quad \mathrm{BR}(H_3 \rightarrow A_2 A_2) = 79.0\%\, , \\ 
&\text{\underline{Benchmark H2:}}\quad M_{H_3} = 741.53~\mathrm{GeV}, \quad  \mathrm{BR}(H_3 \rightarrow H^+ W^-) = 22.3 \%\, , \\
&\text{\underline{Benchmark H3:}}\quad M_{H_3} = 724.46~\mathrm{GeV}, \quad \mathrm{BR}(H_3 \rightarrow A_2 A_2) = 35.6\%\, , \\
&\text{\underline{Benchmark H4:}}\quad M_{H_3} = 705.62~\mathrm{GeV}, \quad \mathrm{BR}(H_3 \rightarrow A_2 \mathrm{Z^0}) = 33.9\%\, , \\
&\text{\underline{Benchmark H5:}}\quad M_{H_3} = 626.92~\mathrm{GeV}, \quad \mathrm{BR}(H_3 \rightarrow H^+ H^-) = 50.3\%\, , \\ 
&\text{\underline{Benchmark H6:}}\quad M_{H_3} = 439.59~\mathrm{GeV}, \quad \mathrm{BR}(H_3 \rightarrow A_2 \mathrm{Z^0}) = 66.9\%\, ,
\end{aligned}
\end{equation}
one can deduce topologies that are distinct from those of the $H_2$ scalar. While a number of interesting channels can be derived from here, one may disregard the benchmarks H1 and H3 as the corresponding decay channel $H_3 \rightarrow A_2 A_2$ has already been studied in dedicated experimental analyses \cite{ATLAS:2021hbr,ATLAS:2021ldb,CMS:2020ffa}. Additionally, we note that for the benchmarks H4 and H6 there have already been searches performed by ATLAS \cite{ATLAS:2020gxx} focusing on the inverse channel, $A\rightarrow H\mathrm{Z^0}$. Our current work represents a novel analysis of $H_3$ production in the benchmarks scenarios H2 and H5 whose final-state topologies, to the best of our knowledge, have not yet been investigated by LHC experiments. Indeed, the corresponding decays are of great interest as they allow to simultaneously probe the masses and/or couplings of both the neutral and charged scalars, with the latter decaying before reaching the detector. For this purpose, one should consider the largest corresponding BRs which, for the benchmark H2, read as
\begin{equation}\nonumber
    \mathrm{BR}(H^+ \rightarrow A_2 W^+) = 72.6\% \,, \quad \mathrm{BR}(A_2 \rightarrow c\bar{c}) = 97.3\% \,.
\end{equation}

\begin{figure}[htb!]
    \captionsetup{justification=raggedright,singlelinecheck=false}
    \centering
    \includegraphics[width=0.5\textwidth]{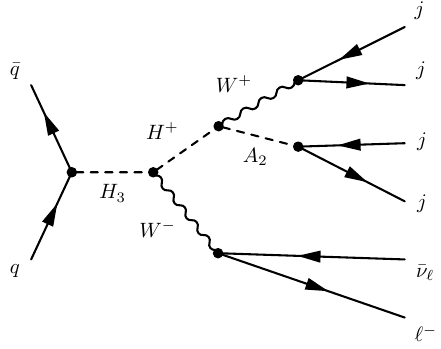}
    \caption{The LO Feynman diagrams for production of the single $H_3$ CP-even scalar. Here, $q$ denotes the SM quarks and $j$ stands for a jet originating from the physical quarks of first and second generations $u,s,d,c$ and their anti-particles. While not explicitly shown here, the gluon-gluon fusion contribution may play an important role in the cross section and hence has been included in the numerical analysis.}
    \label{fig:H3_singleproduction}
\end{figure}

Combining this information, the proposed topology is shown in Fig.~\ref{fig:H3_singleproduction}. As one can notice, this results in a rich final state featuring four jets originating from quarks, one muon and a neutrino which acts as missing transverse energy (MET) in the detector. Besides, note that three of the new scalars contribute in the internal propagators as virtual states within the same topology, thus, opening up a new possibility to constrain the charged, the CP-even and the CP-odd sectors with a single process.

In order to ensure that $H_3$ and $H^\pm $ are produced on-shell, one requires the following mass hierarchy
\begin{equation}\label{eq:mass_hierachy}
M_{H_3} > M_{H^+} + M_{W} \,, \quad M_{H^+} > M_{A_2} + M_{W} \,.
\end{equation}
For the benchmark scenario H2, the condition \eqref{eq:mass_hierachy} is satisfied, with $M_{H^+} = 448.52~\mathrm{GeV}$ and $M_{A_2} = 188.01~\mathrm{GeV}$. Based on the ATLAS analysis \cite{ATLAS:2018pvw}, for this type of topology, the main backgrounds include $t\bar{t}+$jets, $V+$jets (with $V = W,\mathrm{Z^0}$) and single top production. An additional sizeable contribution from fakes (events with either jet or photon that are mistagged as a lepton) and non-prompt leptons (leptons that originate from decays from the hadronized quarks in the jets) is also relevant. For the benchmark H5, the neutral scalar would decay into a pair of charged Higgs bosons. However, based on the available branching ratios, the dominant decays would proceed into a $c \bar{s}$ pair, with $\mathrm{BR}(H^+ \rightarrow c \bar{s}) = 92.8\%$, resulting in an undesirable fully hadronic final state.

\subsection{CP-odd neutral scalars}\label{subsec:CPodd}

In addition to the CP-even states discussed above, our framework features two heavy CP-odd fields, $A_2$ and $A_3$ (with the mass hierarchy of $m_{A_2}<m_{A_3}$) whose signal and background topologies will also be analysed in this work. Again, here we focus on specific topologies suggested by $A_{2,3}$ decay modes with the largest BRs. Starting with the lightest one, $A_2$, the corresponding masses and the dominant BRs for the benchmark scenarios from H1 to H6 read
\begin{equation}\label{eq:BRS_3}\nonumber
\begin{aligned}
&\text{\underline{Benchmark H1:}}\quad M_{A_2} = 314.99~\mathrm{GeV}\,, 
\quad \mathrm{BR}(A_2 \rightarrow c\bar{c}) =  86.5\%\, , \\ 
&\text{\underline{Benchmark H2:}}\quad M_{A_2} = 159.20~\mathrm{GeV}\,, 
\quad \mathrm{BR}(A_2 \rightarrow c \bar{c}) = 97.3 \%\, , \\
&\text{\underline{Benchmark H3:}}\quad M_{A_2} = 205.41~\mathrm{GeV}\,, 
\quad \mathrm{BR}(A_2 \rightarrow c \bar{c}) = 97.3\%\, , \\
&\text{\underline{Benchmark H4:}}\quad M_{A_2} = 188.01~\mathrm{GeV}\,, 
\quad \mathrm{BR}(A_2 \rightarrow c \bar{c}) = 97.3\%\, , \\ 
&\text{\underline{Benchmark H5:}}\quad M_{A_2} = 177.49~\mathrm{GeV}\,, 
\quad \mathrm{BR}(A_2 \rightarrow c \bar{c}) = 96.9\%\, , \\ 
&\text{\underline{Benchmark H6:}}\quad M_{A_2} = 217.36~\mathrm{GeV}\,, 
\quad \mathrm{BR}(A_2 \rightarrow c \bar{c}) = 96.7\%\, . 
\end{aligned}
\end{equation}
We notice that for these scenarios that we have picked, $A_2$ almost always decays into $c\bar c$ pair. So far, most searches at the LHC have focused on the lepton channel (see Tab.~\ref{tab:CMS_ATLAS} and references therein) and mostly on the low-mass domain \cite{ATLAS:2021hbr,ATLAS:2021ldb}, $m_A < 62~\mathrm{GeV}$, i.e.~well below the $A_2$ masses in the above benchmarks. Please do note that the pattern of high BRs to a pair of $c$ quarks should not be interpreted as a feature of the model. Indeed, there are valid parameter points where such decays are not dominant.

\begin{figure}[htb!]
    \captionsetup{justification=raggedright,singlelinecheck=false}
    \centering
    \includegraphics[width=0.39\textwidth]{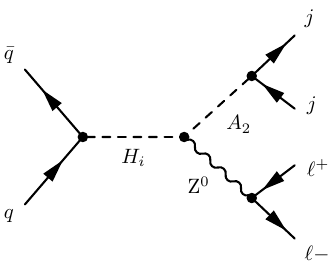}
    \includegraphics[width=0.46\textwidth]{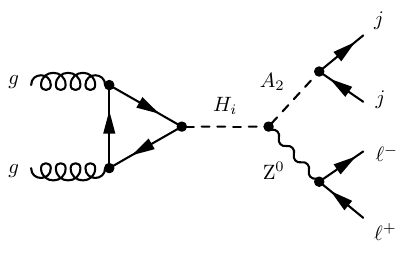} \\
    \includegraphics[width=0.46\textwidth]{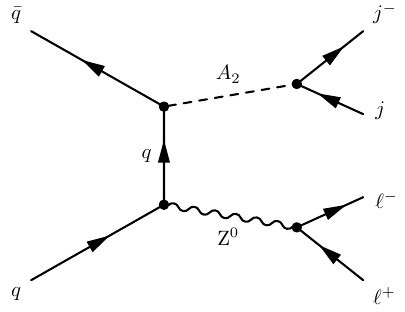}
	\caption{
	The LO Feynman diagrams for production of the single $A_2$ CP-odd scalar in the $s$-channel (top two diagrams) and the $t$-channel (bottom diagram) processes. Here, $q$ denotes the SM quarks, $g$ corresponds to a gluon, and $H_{i=1,2,3}$ -- to CP-even scalars, while $j$ stands for a jet originating from the physical quarks of first and second generations $u,s,d,c$ (and their anti-particles) and $\ell = e,\mu$ denotes the first and second generation charged leptons (and their anti-particles).}
    \label{fig:A2_singleproduction}
\end{figure}

\begin{figure}[htb!]
    \captionsetup{justification=raggedright,singlelinecheck=false}
    \centering
    \includegraphics[width=0.39\textwidth]{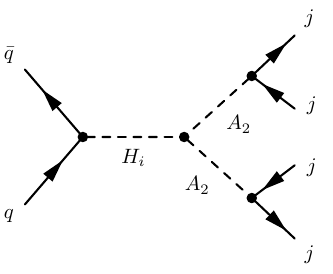}
    \includegraphics[width=0.46\textwidth]{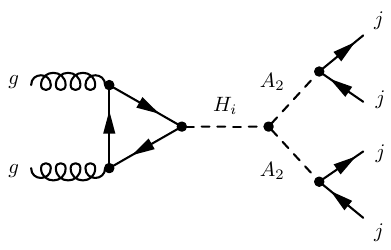} \\
    \includegraphics[width=0.46\textwidth]{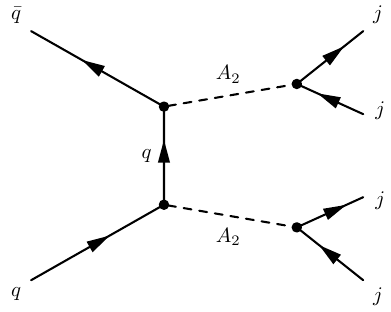}
	\caption{The LO Feynman diagrams for pair-production of the $A_2$ CP-even scalars in the $s$-channel (top two diagrams) and the $t$-channel (bottom diagram) processes. Here, $q$ denotes the SM quarks, $g$ corresponds to a gluon, and $H_{i=1,2,3}$ -- to CP-even scalars, while $j$ stands for a jet originating from the physical quarks of first and second generations $u,s,d,c$ and their anti-particles.}
    \label{fig:A2_pairproduction}
\end{figure}

In what follows, we would like to consider the topologies involving $A_2$ decays into light jets. In particular, we consider very similar topologies to those already shown in Figs.~\ref{fig:H2_singleproduction} and \ref{fig:H2_pairproduction}, with the replacement $H_2 \rightarrow A_2$. Various LO diagrams that contribute to such a signal are shown in Figs.~\ref{fig:A2_singleproduction} and \ref{fig:A2_pairproduction}. Since the final states are identical to those of $H_2$, the main irreducible backgrounds remain the same as for the $H_2$ topologies considered above.

\begin{figure}[htb!]
    \captionsetup{justification=raggedright,singlelinecheck=false}
    \centering
    \includegraphics[width=0.5\textwidth]{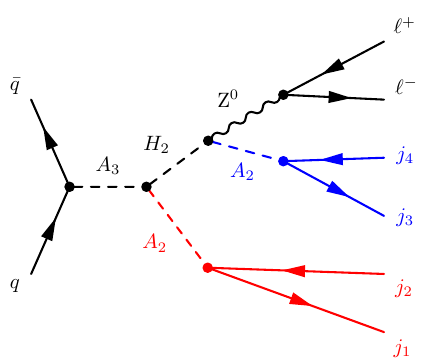}
    \caption{The LO Feynman diagrams for production of the single $A_3$ CP-odd scalar. Here, $q$ denotes the SM quarks, $j$ stands for a jet originating from the physical quarks of first and second generations $u,s,d,c$ (and their anti-particles) and $\ell = e,\mu$ denotes the first and second generation charged leptons (and their anti-particles). While not explicitly shown here, the gluon-gluon fusion contribution may play an important role in the cross section and hence has been included in the numerical analysis.}
    \label{fig:A3_singleproduction}
\end{figure}

For the heavier CP-odd state, $A_3$, the dominant BRs read as
\begin{equation}\label{eq:BRS_4}\nonumber
\begin{aligned}
&\text{\underline{Benchmark H1:}}\quad M_{A_3} = 955.09~\mathrm{GeV}\,, 
\quad \mathrm{BR}(A_3 \rightarrow A_2 H_2) =  24.5\%\, , \\
&\text{\underline{Benchmark H2:}}\quad M_{A_3} = 566.75~\mathrm{GeV}\,, 
\quad \mathrm{BR}(A_3 \rightarrow H_2 \mathrm{Z^0}) = 44.4 \%\, , \\  
&\text{\underline{Benchmark H3:}}\quad M_{A_3} = 707.60~\mathrm{GeV}\,, 
\quad \mathrm{BR}(A_3 \rightarrow H^+ W^-) = 36.1\%\, , \\
&\text{\underline{Benchmark H4:}}\quad M_{A_3} = 611.07~\mathrm{GeV}\,, 
\quad \mathrm{BR}(A_3 \rightarrow H_2 \mathrm{Z^0}) = 38.78\%\, , \\ 
&\text{\underline{Benchmark H5:}}\quad M_{A_3} = 683.60~\mathrm{GeV}\,, 
\quad \mathrm{BR}(A_3 \rightarrow H^+ W^-) = 27.1\%\, , \\ 
&\text{\underline{Benchmark H6:}}\quad M_{A_3} = 772.32~\mathrm{GeV}\,, 
\quad \mathrm{BR}(A_3 \rightarrow A_2 H_2) = 59.6\%\, .
\end{aligned}
\end{equation}
Combining this information with that of Tab.~\ref{tab:CMS_ATLAS}, we note that the preferred channels in the search for $A_3$ would rely on $A_3\rightarrow A_2 H_2$ (in the benchmark scenarios H1 and H6) or $A_3\rightarrow H^+ W^-$ (in the benchmark scenarios H3 and H5) decays. The remaining benchmarks H2 and H4 result in topologies that have been probed already by ATLAS Collaboration in \cite{ATLAS:2020gxx}. Both dominant $A_3$ decay modes in H1/H6 and H3/H5 are interesting since the vast majority of analyses performed at the LHC so far typically do not consider decay channels that involve multiple BSM particles in the internal propagators (the existing examples are listed in Tab.~\ref{tab:CMS_ATLAS}), which results in considerably weaker constraints on these decay modes. 

In our analysis, we first focus on the decay $A_3\rightarrow A_2 H_2$ which is dominant in the benchmark scenarios H1 and H6. Then, combining with the information on the BRs for $H_2$ and $A_2$ shown above, we deduce that the optimal final state in this case would be $4j+2\ell$, where we assume a $\mathrm{Z^0}$ boson decaying into a pair of leptons. The LO Feynman diagram for such a topology is shown in Fig.~\ref{fig:A3_singleproduction}, with the background being dominated by $\mathrm{Z^0}+$jets, $t\bar{t}$ and di-boson production processes.

\subsection{Charged scalars}\label{subsec:ChargedScalars}

\begin{figure}[htb!]
    \captionsetup{justification=raggedright,singlelinecheck=false}
    \centering
    \includegraphics[width=0.40\textwidth]{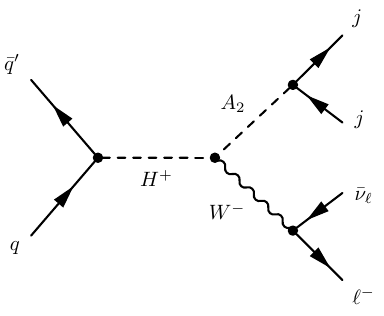}
    \caption{The LO Feynman diagrams for production of the single $H^+$ charged Higgs scalar. Here, $q$ denotes the SM quarks and $j$ stands for a jet originating from the physical quarks of first and second generations $u,s,d,c$ and their anti-particles.}
    \label{Hcharged_production}
\end{figure}

For each of the six considered benchmark points, the masses and the largest BRs for the charged scalar $H^\pm$ read as
\begin{equation}\label{eq:BRS_5}\nonumber
\begin{aligned}
&\text{\underline{Benchmark H1:}}\quad M_{H^\pm} = 566.40~\mathrm{GeV} \,, 
\quad \mathrm{BR}(H^+ \rightarrow A_2 W^+) =  93.6\%\, , \\ 
&\text{\underline{Benchmark H2:}}\quad M_{H^\pm} = 411.94~\mathrm{GeV} \,, 
\quad \mathrm{BR}(H^+ \rightarrow A_2 W^+) = 72.6 \%\, , \\ 
&\text{\underline{Benchmark H3:}}\quad M_{H^\pm} = 524.70~\mathrm{GeV} \,, 
\quad \mathrm{BR}(H^+ \rightarrow A_2 W^+) = 91.0\%\, , \\
&\text{\underline{Benchmark H4:}}\quad M_{H^\pm} = 448.52~\mathrm{GeV} \,, 
\quad \mathrm{BR}(H^+ \rightarrow A_2 W^+) = 88.8\%\, , \\ 
&\text{\underline{Benchmark H5:}}\quad M_{H^\pm} = 156.89~\mathrm{GeV} \,, 
\quad \mathrm{BR}(H^+ \rightarrow c \bar{s}) = 92.8\%\, , \\
&\text{\underline{Benchmark H6:}}\quad M_{H^\pm} = 330.88~\mathrm{GeV} \,, 
\quad \mathrm{BR}(H^+ \rightarrow A_2 W^+) = 63.8\%\, .
\end{aligned}
\end{equation}
Here, one notices that there are two interesting distinct scenarios that can be probed at the LHC. In particular, the benchmarks scenarios H1 to H4 and H6 result in the following process: $pp\rightarrow H^+ \rightarrow A_2 W^+ \rightarrow j j \ell^+ \nu_\ell$. Note that the majority of the experimental studies focusing on the charged Higgs boson search assume its production via the $tbH^+$ coupling. The more interesting and novel channels that can be considered are those where $H^+$ is produced via lighter quarks as shown by a LO diagram in Fig.~\ref{Hcharged_production}. Additionally, for the benchmark H5, the $H^+$ coupling to light quarks, $c\bar{s}$, is dominant. Indeed, as noted throughout this work, a substantial part of the preferred decay channels of the new scalars tend to be those involving decays to light quarks, which are not well constrained by collider measurements so far. The main irreducible background is expected to be dominated by $\mathrm{W}+$jets, di-boson and $t\bar{t}$ production processes.


\subsection{Production cross sections}\label{subsec:cross-sections}

For the topologies introduced above, we have computed the corresponding production cross sections for proton-proton collisions at the centre-of-mass energy of $\sqrt{s}=14~\mathrm{TeV}$ using \texttt{MadGraph}. These results are listed in Tab.~\ref{tab:cross_sections_v1}.
\begin{table}[htb!]
\centering
\captionsetup{justification=raggedright,singlelinecheck=false}
	\begin{tabular}{c|c|c|c|c}
			& Cross section & Events at run-III & Events at HL-LHC & Benchmarks \\ \hline
			Figure 1 ($H_2$ single prod.) &	
			\makecell{
				\begin{math}
				\sigma =  0.172~\mathrm{fb}
				\end{math}
			}
			&
			51
			&
			515
			&
			H5
		    \\
		    \hline
			Figure 2 ($H_2$ double prod.)  &
			\makecell{
				\begin{math}
				\sigma =  0.350~\mathrm{fb}
				\end{math}
			}
			&
			105
			&
			1050
			&
			H5
			\\
			\hline 
			Figure 3 ($H_3$ single prod.) &
			\makecell{
				\begin{math}
				\sigma = 0.194~\mathrm{fb}
				\end{math}
			}
			&
			58
			&
			582
			&
			H2
			\\
			\hline 
			Figure 4 ($A_2$ double prod.)  &
			\makecell{
				\begin{math}
				\sigma = 24.40~\mathrm{fb}
				\end{math}
			}
			&
			7320
			&
			73200
			&
			H4
			\\
			\hline      
			Figure 5 ($A_2$ single prod.)  &
			\makecell{
				\begin{math}
				\sigma = 7.71~\mathrm{fb}
				\end{math}
			}
			&
			2313
			&
			23130
			&
			H4
			\\
			\hline
			Figure 6 ($A_3$ single prod.)  &
			\makecell{
				\begin{math}
				\sigma = 0.0594~\mathrm{fb}
				\end{math}
			}
			&
			17
			&
			178
			&
			H1
			\\
			\hline
			Figure 7 ($H^+$ single prod.)  &
			\makecell{
				\begin{math}
				\sigma = 34.5~\mathrm{fb}
				\end{math}
			}
			&
			10350
			&
			103500
			&
			H1
			\\
			\hline
	\end{tabular}
	\captionsetup{justification=raggedright,singlelinecheck=false}
	\caption{The predicted total cross sections $\sigma$ (in fb) for each of the topologies introduced in the previous section. The calculation is made at the LO level. Additionally, we present the expected numbers of events found as $N=\sigma\mathcal{L}$ at both the LHC Run-III (at luminosity of $\mathcal{L}=300~\mathrm{fb^{-1}}$) and the HL-LHC (at luminosity of $\mathcal{L}=3000~\mathrm{fb^{-1}}$) phases.}
	\label{tab:cross_sections_v1}
\end{table}
One can readily observe that every single topology for the selected benchmarks shown in Tab.~\ref{tab:cross_sections_v1} can potentially be probed both at the LHC Run-III and at the HL-LHC. Note that all considered signal processes are within the reach of the current data for the Run-II luminosity of $\mathcal{L}=139~\mathrm{fb^{-1}}$. It is then possible to look for these signals without having to wait for new batches of data.

In fact, we can deduce from such a simple calculation that the single $A_3$ production process is the most challenging for future searches, with the smallest amount of expected events, \textit{i.e.}~$N=17$ at the LHC Run-III and $N=178$ at the HL-LHC. On the other hand, production of such a scalar leads to a very interesting final state, with four jets and two charged leptons. In this case, the jets originate from $A_2$, whose mass in the selected benchmark scenarios varies between 150 and 400 GeV. As noted in \cite{ATLAS:2018pvw}, the most relevant background processes for this type of processes are due to $\mathrm{Z^0}+$jets and $t\bar{t}$ final states, while the other contributions are subleading. Besides requiring that the invariant mass of the lepton pair is close to the mass of the $\mathrm{Z^0}$ boson, one also requires that the invariant mass of two jet pairs is in a vicinity of the $A_2$ mass (see \cref{fig:A3_singleproduction}), thus enabling one to reduce the dominant $\mathrm{Z^0}+$jets background. In addition, the fact that the signal features a high-multiplicity of jets may also help in reducing the dominant background. In this work, we perform a detailed analysis of the single $A_3$ production process shown in \cref{fig:A3_singleproduction}. A similar treatment of other topologies is left for a future work.

\section{Single $A_3$ production: feature selection and cuts}\label{sec:Collider}

The events are generated for proton-proton collisions and at the maximal LHC centre-of-mass energy of $\sqrt{s} = 14$ TeV for both the signal processes illustrated in \cref{fig:A3_singleproduction} and the background processes discussed in \cite{ATLAS:2018pvw}. In practical calculations, we adopt the QCD collinear parton distribution functions in the proton, the strong coupling constant and their evolution from the NNPDF2.3 analysis \cite{Ball:2013hta}. 

Our study starts with the model implementation in \texttt{SARAH} \cite{Staub:2013tta}, where the Lagrangian in the gauge basis has been implemented. From \texttt{SARAH}, we compute all interaction vertices in the mass basis and export them to \texttt{MadGraph} (MG5) \cite{Alwall:2014hca} which is used to compute the matrix elements for signal, $\mathrm{Z^0}+$jets and di-boson background topologies at the LO level. The cross sections for $t\bar{t}$ and single-$t$ background processes have been adopted from the literature instead. In particular, the $t\bar{t}$ production cross section is normalized to the theoretical calculation at next-to-next-to leading order (NNLO) in perturbative QCD. The mass dependence of the $t\bar{t}$ cross section can be parametrized by \cite{Czakon:2013goa}
\begin{equation}\label{eq:cross_section_ttbar}
\sigma(m_t) = \sigma(m_\mathrm{ref}) \qty(\frac{m_\mathrm{ref}}{m_t})^4 \qty[1 + a_1 \qty(\frac{m_t - m_\mathrm{ref}}{m_\mathrm{ref}}) + a_2 \qty(\frac{m_t - m_\mathrm{ref}}{m_\mathrm{ref}})^2] \,,
\end{equation}
where $m_t$ is the top-quark mass chosen in the Monte Carlo generator, $m_\mathrm{ref}$ is a reference mass scale, $\sigma(m_\mathrm{ref})$ is the cross section at that scale, and $a_{1,2}$ are the fitting parameters. In \texttt{MadGraph} we set the top mass at 173 GeV. For the values of the fitting parameters known from \cite{ttbarnnlo}, the total $t\bar{t}$ cross section is $\sigma(m_t) = 985.543~\mathrm{pb}$. Similarly, the next-to-leading order (NLO) corrections to the single-top production cross section are also included, providing us with: $\sigma(t-\mathrm{channel}) = 248.05~\mathrm{pb}$, $\sigma(s-\mathrm{channel}) = 11.39~\mathrm{pb}$ and $\sigma(Wt-\mathrm{channel}) = 83.56~\mathrm{pb}$ (see \cite{SingleTopnlo} for further details).

For the signal topology, we separately generate samples of 600k events for each of the six benchmark scenarios described above, corresponding to different values of scalar bosons' masses and BRs. For this purpose, a total of 3M events for $\mathrm{Z^0}+$jets, 1M events for $t\bar{t}$ and 300k events for other processes (with di-boson, single top and $t\bar{t}$ plus vector boson final states) are generated. For the SM background one has used the default UFO codes and parameter cards available in \texttt{MadGraph}. Our samples contain up to four jets, emerging from both light and heavy quarks. For matching and merging of the jets to the original quarks, we rely on the automatic procedures available in \texttt{MadGraph}, which by default considers the kT-MLM matching scheme \cite{Mangano:2006rw}. For jet tagging, we employ the following strategy: for each event, we pick all existing jets (which in practical terms involves getting all entries inside the \texttt{Jet} class on \texttt{Delphes}) and impose a b-tagging method, such that then we can divide into a list of jets tagged as originating from b-quarks and another list of jets not tagged as such, which we identify as a list of light jets. Since in the signal topology, we do not consider the case where  the $A_2$ fields decay into $b$ quarks, the list of $b$-jets is discarded. We also do not apply any tau tagging algorithm in the selection of jets.

Since the LHC is a hadron collider, production of scalars via gluon-gluon fusion is also of relevance. While couplings between scalars and gluons are not present at tree-level, they appear at one-loop level via triangle quark loops. This contribution is determined by considering the effective $gg\Phi$ coupling, where $\Phi$ is a scalar and $g$ is the gluon. The computation of this effective coupling is done via \texttt{SPheno} \cite{Porod:2011nf} and has subsequently been used as an input in the \texttt{MadGraph} parameter card. Hadronization and showering of the final-state particles have been performed with \texttt{Pythia8} \cite{Sjostrand:2014zea}. Then, the fast simulation of the ATLAS detector is conducted with \texttt{Delphes} \cite{deFavereau:2013fsa}, from where we extract the relevant kinematic information about the final-state particles using \texttt{ROOT} \cite{Brun:1997pa}.
\begin{table}[htb!]    
	\centering
	\captionsetup{justification=raggedright,singlelinecheck=false}
	\resizebox{1.0\textwidth}{!}{\begin{tabular}{|c|c|c|c|}
			\toprule
			\hline
			& Dimension-full & \multicolumn{2}{|c|}{Dimensionless} \\
			\hline
			\hline
			\makecell{Signal}  & \makecell{$p_T(\ell^-)$, $p_T(\ell^+)$, $p_T(j_a)$, $E(\ell^-)$,\\
			$E(\ell^+)$, $E(j_a)$, $M(\mathrm{Z^0})$, $M(A_{2,1})$, $M(A_{2,2})$,\\ $M(H_2)$, $p_T(\mathrm{Z^0})$, $p_T(A_{2,1})$,  $p_T(A_{2,2})$, $p_T(H_2)$, \\ $M(A_3)$, $p_T(A_3)$} & 
			\makecell{
				$\cos(\theta_{j_a j_b})$, $\cos(\theta_{j_a \ell^+})$, \\
				$\cos(\theta_{j_a \ell^-})$,$\cos(\theta_{\ell^+ \ell^-})$,$\eta(\ell^+)$, $\eta(\ell^-)$,\\
				$\eta(j_a)$, $\phi(\ell^+)$, $\phi(\ell^-)$, $\phi(j_a)$, \\  $\eta(\mathrm{Z^0})$, $\phi(\mathrm{Z^0})$, $\eta(j_a,j_b)$, $\phi(j_a,j_b)$, \\ $\eta(A_3)$, $\phi(A_3)$, $\eta(A_{2,1})$, $\eta(A_{2,2})$ \\ $\phi(A_{2,1})$,$\phi(A_{2,2})$, $\eta(H_2)$, $\phi(H_2)$, \\
				$\cos(\theta_{A_{2,1} A_{2,2}})$, $\cos(\theta_{A_{2,1} H_2})$, \\ $\cos(\theta_{A_{2,1} \mathrm{Z^0}})$, $\cos(\theta_{A_{2,2} \mathrm{Z^0}})$, \\ $\cos(\theta_{H_{2} \mathrm{Z^0}})$} & 
			\makecell{$\Delta R(j_a, j_b)$, $\Delta R(\ell^+, \ell^-)$, $\Delta R(j_a, \ell^+)$, \\
				$\Delta R(j_a, \ell^-)$, $\Delta \Phi(j_a, j_b)$, $\Delta \Phi(\ell^+, \ell^-)$, \\
				$\Delta \Phi(j_a, \ell^+)$, $\Delta \Phi(j_b, \ell^-)$, $\Delta \Theta(j_a, j_b)$, \\
				$\Delta \Theta(\ell^+, \ell^-)$, $\Delta \Theta(j_a, \ell^+)$, $\Delta \Theta(\j_a, \ell^-)$, \\
				$\Delta\Phi (A_{2,1}, A_{2,2})$, $\Delta\Phi (A_{2,1}, H_{2})$, $\Delta\Phi (A_{2,2}, H_{2})$, \\
				$\Delta\Phi(A_{2,1},\mathrm{Z^0})$, $\Delta\Phi(A_{2,2},\mathrm{Z^0})$,
				$\Delta\Phi(H_{2},\mathrm{Z^0})$, \\
				$\Delta\Theta (A_{2,1}, A_{2,2})$, $\Delta\Theta (A_{2,1}, H_{2})$, $\Delta\Theta (A_{2,2}, H_{2})$, \\
				$\Delta\Theta(A_{2,1},\mathrm{Z^0})$, $\Delta\Theta(A_{2,2},\mathrm{Z^0})$,
				$\Delta\Theta(H_{2},\mathrm{Z^0})$, \\				
				$\Delta R (A_{2,1}, A_{2,2})$, $\Delta R (A_{2,1}, H_{2})$, $\Delta R (A_{2,2}, H_{2})$, \\
				$\Delta R(A_{2,1},\mathrm{Z^0})$, $\Delta R(A_{2,2},\mathrm{Z^0})$,
				$\Delta R(H_{2},\mathrm{Z^0})$}\\
			\hline
			\hline
	\end{tabular}}
	\caption{Angular and kinematic observables selected for our study of the signal topology shown in Fig.~\ref{fig:A3_singleproduction}. All variables are computed in the laboratory frame. To simplify the presentation, we define $(j_a,j_b)$ to indicate different combinations of jets. Reconstructed variables involve various final states, which are shown in parentheses. Variables that involve $A_3$ are reconstructed using all final states ($A_3 = \sum_{i=1}^4 j^\mu_i + \ell^{\mu,+} + \ell^{\mu,-}$).}
	\label{tab:vars}
\end{table}

The input data is composed of normalised multidimensional distributions based on kinematical and angular observables reconstructed from final state particles. The relevant kinematical variables are summarized in Table~\ref{tab:vars}. Note that, for our signal, the final state is comprised of four light-jets meaning that, in general, one can not distinguish between the jets originating from the first $A_2$ state and those from the second one. However, through the use of kinematic properties, one can correctly match the jets with a particular $A_2$ they originate from. For instance, if all the jets are correctly matched to their original $A_2$ parent, the reconstructed invariant mass difference between $A_2$ states must be small, within a given threshold $\varepsilon$, i.e.~$\Delta M \equiv M(j_1, j_2) - M(j_3, j_4) < \varepsilon~\mathrm{GeV}$. 

As such, in order to perform the correct matching, we take all the jets in each event and compute $\Delta M$ for every single combination. From those combinations that lead to a $\Delta M$ below the threshold, we select the smallest mass difference. In fact, such a restriction heavily impacts the backgrounds. Indeed, the mass difference between the two pairs of jets will unlikely be small, and as such the backgrounds are expected to lead to arbitrary values of $\Delta M$, heavily reducing the backgrounds, in particular, the $\mathrm{Z^0}+\mathrm{jets}$ one.

Once the jets have been correctly matched to the corresponding parent $A_2$ states, we need to also match one of the jet pairs and the lepton pair to the $H_2$ state. In particular, there must exist a given combination that approaches the decay threshold, $M_{H_2} = M_{A_2} + M_{\mathrm{Z^0}}$. To determine this, we consider the jet combinations that most closely match the mass of $H_2$ to be the one associated with the upper $A_2$ blue whereas the other pair corresponds to the lower $A_2$ red as depicted in Fig.~\ref{fig:A3_singleproduction}. 

\begin{figure}[htb!]
    \centering
    \captionsetup{justification=raggedright,singlelinecheck=false}
	\includegraphics[width=\textwidth]{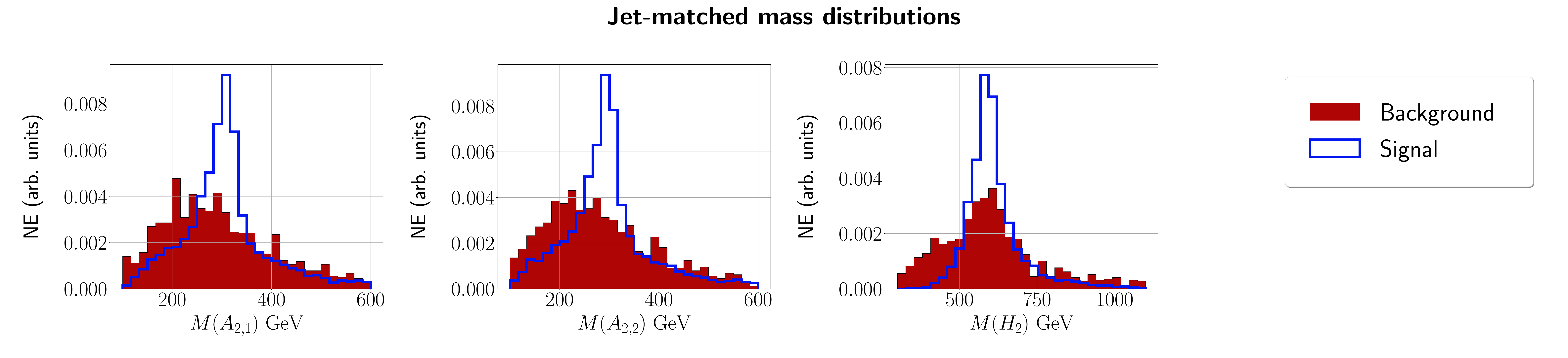}
	\caption{The mass distributions for both $A_2$ and $H_2$ scalars for the benchmark H1. We adopt the notation such that $A_{2,1}$ is the $A_2$ that originates from the $H_2$ parent, whereas $A_{2,2}$ originates from $A_3$. In the $y$-axis, we show the normalized events, and we consider 30 bins for both signal and background components. The mass observables are displayed in units of GeV.}
	\label{fig:Mass_dist} 
\end{figure}

With this technique, we can successfully reconstruct all particles in the chain. Considering $\Delta M < 25~\mathrm{GeV}$, this can be observed in \cref{fig:Mass_dist}, where the mass distributions of both $A_2$ and $H_2$ are presented. One notices a very well defined Breit-Wigner shape, indicating a successful reconstruction. However, such a Breit-Wigner distribution is not perfect with an accumulation of events right below the mass peak (in particular, for the $A_2$ mass plots). This result indicates that the method achieves the full reconstruction leaving only some jet pairs that were incorrectly matched. However, since the peak is well defined, we are confident that the vast majority of combinations are correct. Naturally, using the Monte Carlo history, one can identify all the jets. However, that method can not be extended to real data, whereas using the detailed kinematical information of the jets can be more easily realised in experimental setups.

Besides the constraint on the mass difference, we also impose additional cuts which help to further separate signal and background events. In particular, we consider the following cuts
\begin{equation}\nonumber
	\begin{aligned}
		&p_T(\ell^\pm) > 20 \text{ GeV}, \\
		&\abs{\eta(\ell^\pm)} \leq 2.47,\\
		&p_T(\mathrm{jet}) > 20 \text{ GeV}  \\ 
		&\abs{\eta(\mathrm{jet})} \leq 2.5 \, \rm{and}\\
		&M(\mathrm{jet}) > 15~\mathrm{GeV}.
	\end{aligned}
\end{equation}
Since we have a large number of different kinematic/angular variables, it is expected that some of these do not offer any significant discriminating power. In particular, $\phi$ distributions are typically flat over the entire range of angles and tend to be poor discriminators between the signal and the background. As such, we tested multiple combinations of variables and selected the one giving the best result, in a pre-analysis phase with the Toolkit for Multivariate Data Analysis (\texttt{TMVA}) \cite{Hocker:2007ht} library. From here, we have determined the set of best ten observables that offer a greater discrimination power of the signal over the background.
\begin{figure}[htb!]
    \centering
    \captionsetup{justification=raggedright,singlelinecheck=false}
	\includegraphics[width=\textwidth]{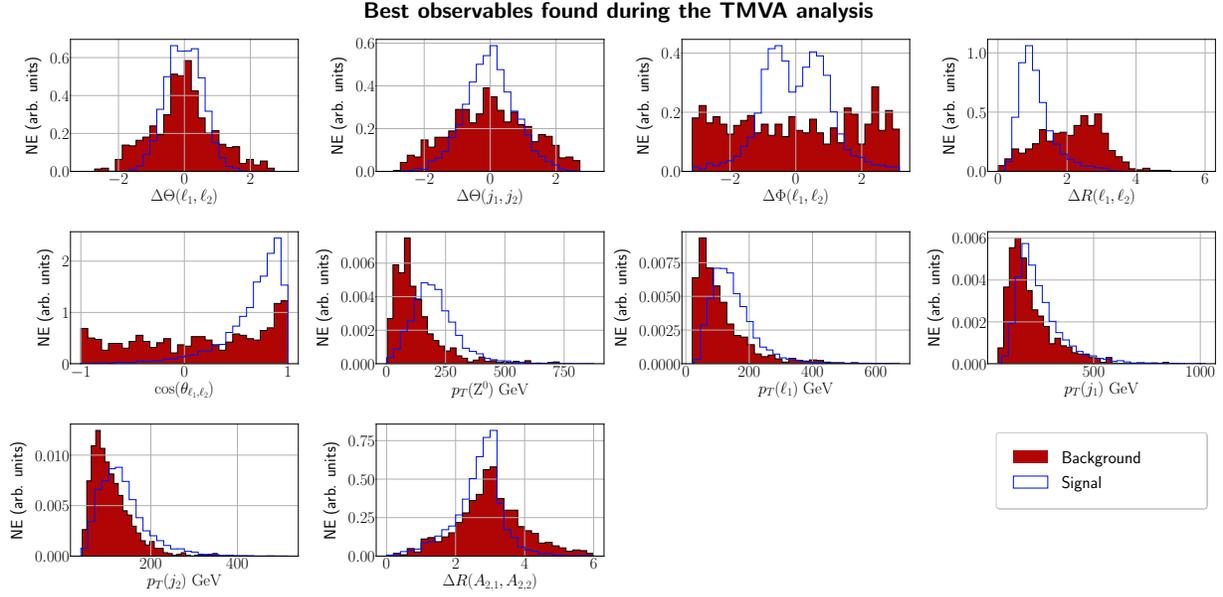}
	\caption{The kinematic and angular observables which were given more importance in the classification by the \texttt{TMVA} toolkit, for the signal sample H1. Reading from left to right and top to bottom, we have $\Delta \Theta (\ell_1, \ell_2)$, $\Delta \Theta (j_1, j_2)$, $\Delta \Phi (\ell_1, \ell_2)$, $\Delta R (\ell_1, \ell_2)$, $\cos(\theta_{\ell_1, \ell_2})$, $p_T(\mathrm{Z^0})$, $p_T (\ell_1)$, $p_T(j_1)$, $p_T(j_2)$ and $\Delta R(A_{2,1}, A_{2,2})$. In $y$-axis, we show normalized events and we consider 30 bins for both the signal and the backgrounds. Dimensionful observables are displayed in units of GeV. The signal histograms are shown in blue and the background ones -- in red.}
	\label{fig:best_observables} 
\end{figure}

\begin{figure}[htb!]
    \centering
    \captionsetup{justification=raggedright,singlelinecheck=false}
	\includegraphics[width=0.9\textwidth]{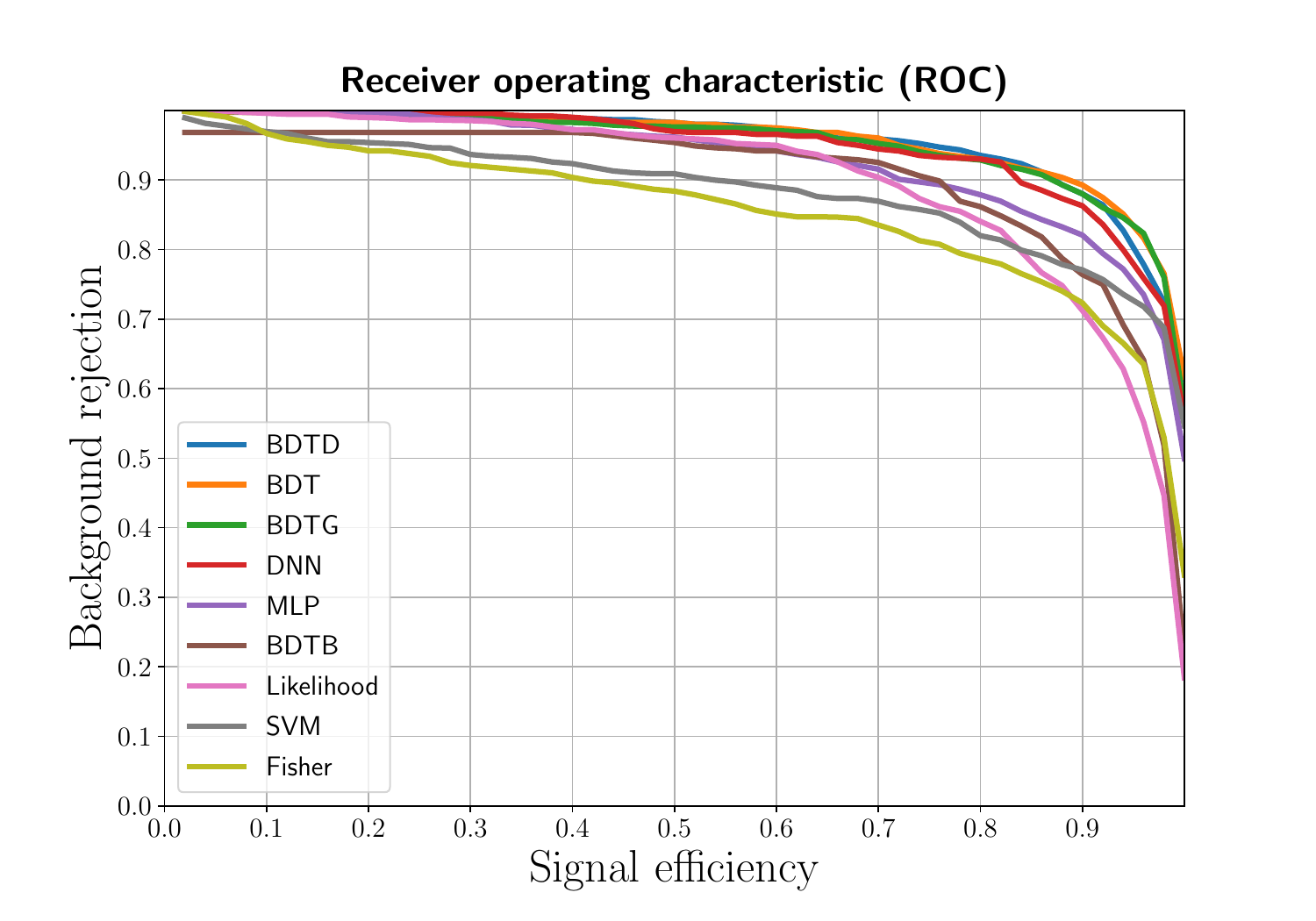}\\
	\caption{ROC curves for each of the 9 methods of classification tests in the \texttt{TMVA} 
	analysis for the H1 benchmark.}
	\label{fig:ROC_curve} 
\end{figure}

\begin{figure}[htb!]
    \centering
    \captionsetup{justification=raggedright,singlelinecheck=false}
	\includegraphics[width=0.85\textwidth]{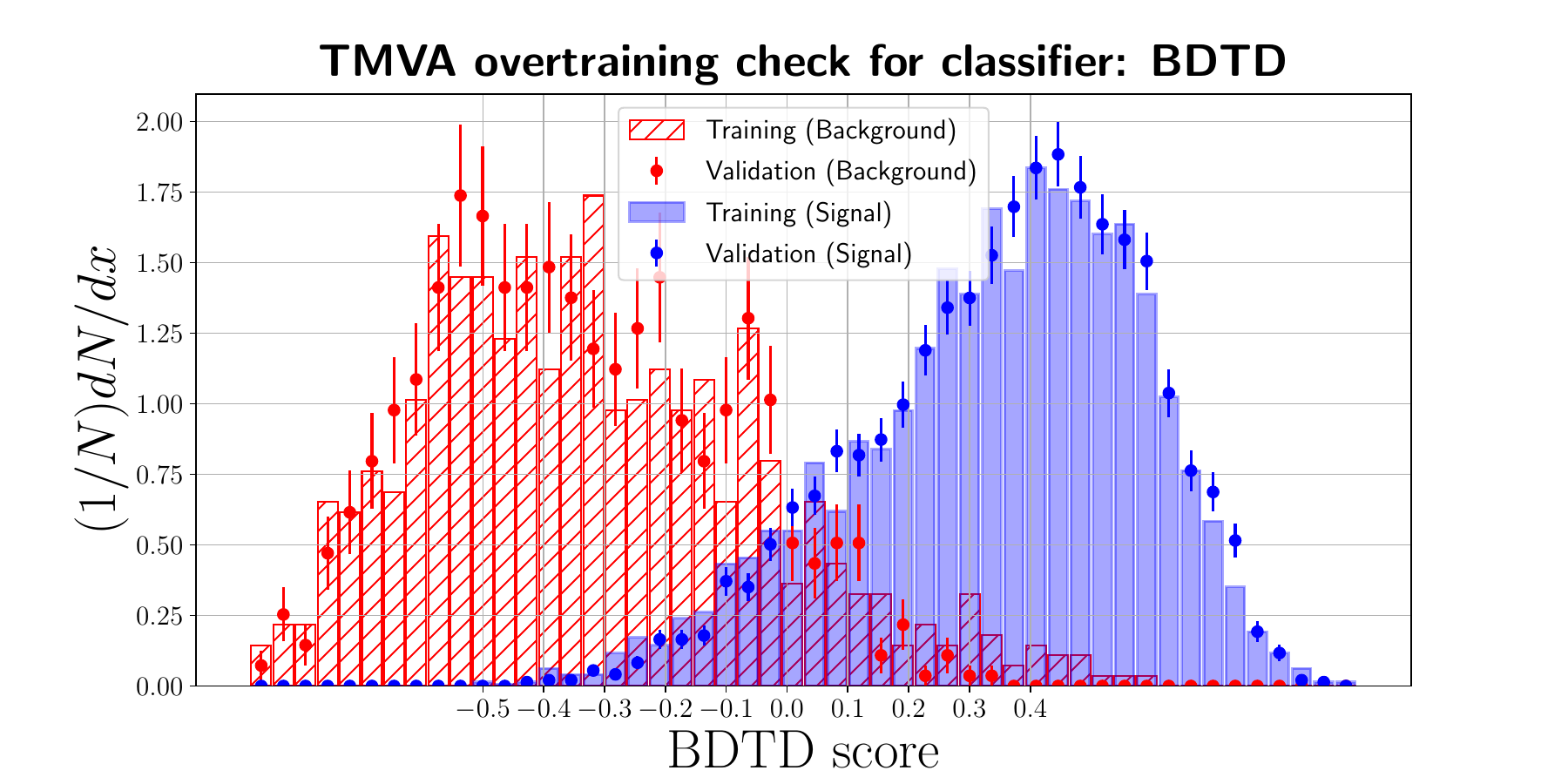}\\
	\includegraphics[width=0.85\textwidth]{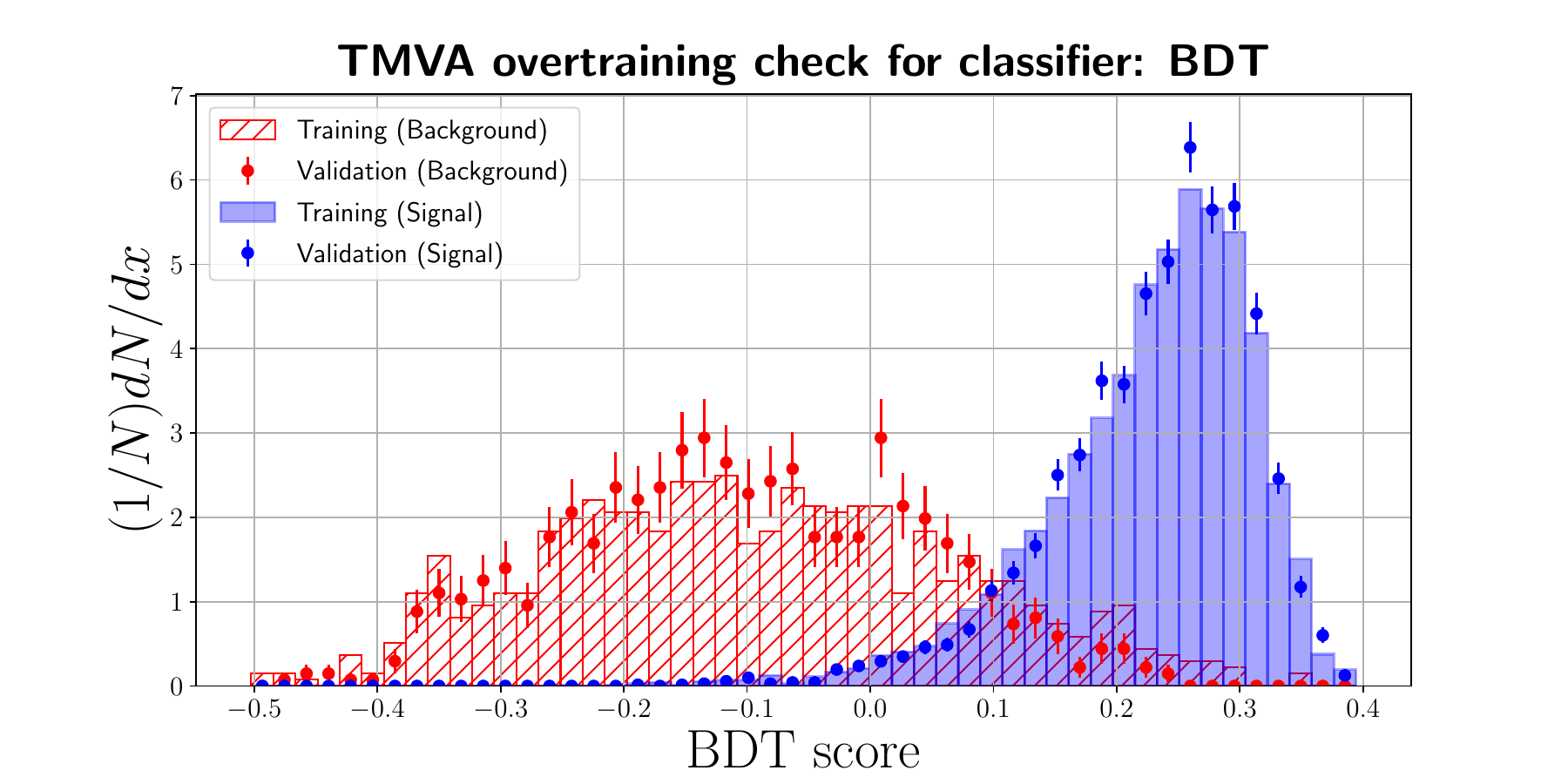}\\
	\includegraphics[width=0.85\textwidth]{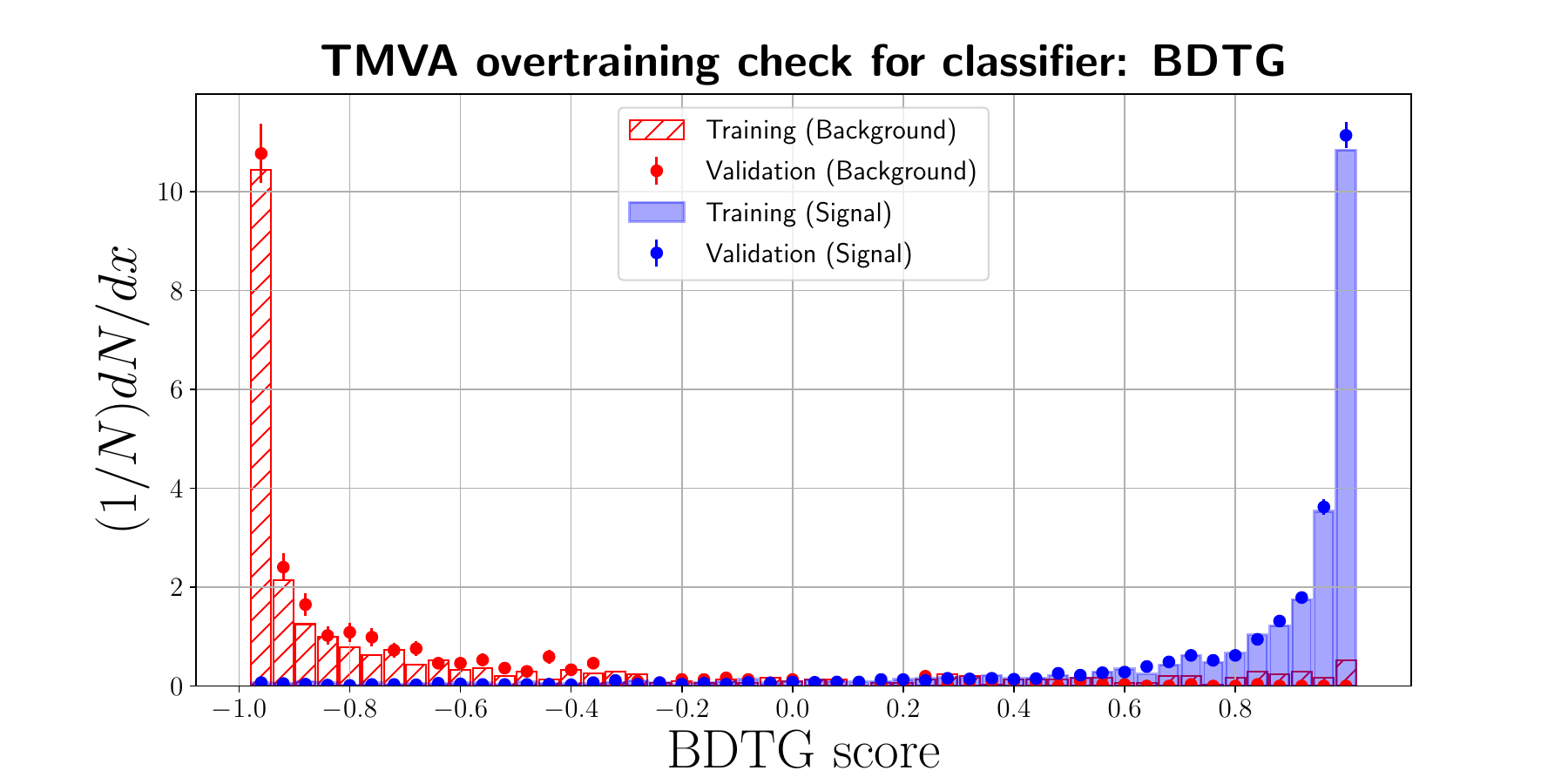}
	\caption{Distributions of the BDTD (top), BDT (middle) and BDTG (bottom) discriminants for the signal (in blue) and the background (in red) corresponding to the H1 benchmark. The results from the training (histograms) and validation (points with error bars) datasets are superimposed.}
	\label{fig:BDT_discr} 
\end{figure}

As a suitable example, let us focus on the sample H1, with masses of the BSM fields fixed as\footnote{For their exact values, see Sec.~\ref{sec:BSM_scalars}.} $M_{H_2} = 600~\mathrm{GeV}$, $M_{H_3} = 900~\mathrm{GeV}$, $M_{A_2} = 300~\mathrm{GeV}$ and $M_{A_3} = 1~\mathrm{TeV}$. In this analysis, we consider 9 different classification tests included in the toolkit, namely the Fisher, Likelihood, BDT, BDTD, BDTG, BDTB, SVM, DNN and MLP methods. We select the top ten observables which correspond to the method that offered the best accuracy in separating the signal from the backgrounds. In our case, we have found that the BDTD method performed the best. The observables that the BDTD method found to offer the best results are presented in Fig.~\ref{fig:best_observables}. Our initial estimates indicate that the best observables are $\Delta \Theta (\ell_1, \ell_2)$, $\Delta \Theta (j_1, j_2)$, $\Delta \Phi (\ell_1, \ell_2)$, $\Delta R (\ell_1, \ell_2)$, $\cos(\theta_{\ell_1, \ell_2})$, $p_T(\mathrm{Z^0})$, $p_T (\ell_1)$, $p_T(j_1)$, $p_T(j_2)$ and $\Delta R(A_{2,1}, A_{2,2})$ including both kinematic and angular variables. Notably, the mass distributions are not advantageous for the signal-background separation, typically under-performing in the classification. While the mass distributions of the backgrounds such as $\mathrm{Z^0}+\mathrm{jets}$ do not peak in the considered regions, the distributions from single-top and $t\bar{t}$ populate those domains of the phase space due to the top quark contribution running in internal propagators. In our analysis we have found that, indeed, angular distributions tend to offer a better separation than the kinematic ones. Additional information is shown in Fig.~\ref{fig:ROC_curve}, where we plot the receiver operating characteristic (ROC) curves for the 9 different methods of classification tests, and in Fig.~\ref{fig:BDT_discr}, where we show the discriminant distributions for the three best performing methods -- BDTD, BDT and BDTG. Note that one can achieve a very good separation between the signal and the backgrounds using only the distributions shown in Fig.~\ref{fig:best_observables}. We also notice that, despite some small fluctuations, the test data, indicated with dots in \cref{fig:BDT_discr}, closely follows the results from the training dataset indicating low overfitting.

This analysis strongly suggests that we can achieve a very good signal-from-background separation. Of course, in order to be able to establish exclusion limits on the BSM particles, one needs to make sure that the cuts involved in the analysis keep a substantial amount of the signal, while significantly reducing the corresponding backgrounds. 

In Tabs.~\ref{tab:cross_sections} and \ref{tab:cross_sections_2}, we present the results for the predicted cross sections before and after imposing the selection criteria, as well indicating the number of expected events at both run-III and the HL phase of the LHC. In Tab.~\ref{tab:cross_sections}, in particular, the jet mass cuts are set at $M(j) > 15~\mathrm{GeV}$ and $\Delta M < 25~\mathrm{GeV}$, whereas in Tab.~\ref{tab:cross_sections_2} the cuts are set to $M(j) > 10~\mathrm{GeV}$ and $\Delta M < 35~\mathrm{GeV}$. First, we notice a massive reduction of the backgrounds. As an example, consider the most dominant one before the cuts, namely, $\mathrm{Z^0}+\mathrm{jets}$. After the selection criteria are imposed, the background is reduced by almost five-six orders of magnitude, from $4.12\times 10^6~\mathrm{fb}$ to $9.64~\mathrm{fb}$ in Tab.~\ref{tab:cross_sections} and to $92.25~\mathrm{fb}$ in Tab.~\ref{tab:cross_sections_2}. The same conclusion holds for the remaining background processes. Indeed, unlike the analysis of \cite{ATLAS:2018pvw}, the $t\bar{t}$ and single top backgrounds now become more important than those from $\mathrm{Z^0}+\mathrm{jets}$. Similar to \cite{ATLAS:2018pvw}, the diboson and $t\bar{t}+V$ backgrounds remain negligible.

Regarding the signal, we note that the more constraining cuts lead to a greater reduction of the cross section, as expected. For example, for the benchmark H1, we see from Tab.~\ref{tab:cross_sections} that the signal is reduced by one order of magnitude, from $0.0594~\mathrm{fb}$ to $0.0064~\mathrm{fb}$, while from Tab.~\ref{tab:cross_sections_2} we have a reduction of the signal by a factor of two. For the benchmark H2, the reduction is much sharper, from $0.16~\mathrm{fb}$ to $0.000699~\mathrm{fb}$ in Tab.~\ref{tab:cross_sections} and to $0.0048~\mathrm{fb}$ in Tab.~\ref{tab:cross_sections_2}. Additionally, we have found that for the benchmarks H2, H3, H4 and H5, the imposed selection criteria are too restrictive and no signal events survive. 

A sharp observed reduction of the signal for the benchmark H6 is related with the masses of the scalar fields in the decay chain. As noted in section~\ref{sec:BSM_scalars}, the masses of scalars in benchmark H1 are larger in comparison to those in benchmark H6. Therefore, the outgoing jets originating from $A_2$ in benchmark H1 are expected to have higher $p_T$'s and masses in comparison to those originating from the $H_2$ scalar. Indeed, greater efficiencies are expected for the larger mass regions. Additionally, this reduction also implies a low number of events at the LHC. For the mass cuts of $M(j) > 15~\mathrm{GeV}$ and $\Delta M < 25~\mathrm{GeV}$, we expect only two events at run-III and 19 events at HL-LHC, for the benchmark H1, whereas for the benchmark H6 we predict only two events surviving the considered cuts at the HL phase. Indeed, for these scenarios, probing such a signal topology may be difficult at run-III, as a low number of events associated with the stringent cuts on the mass of the jets may be overshadowed by systematic/statistical uncertainties present in a real experimental setup. In fact, by relaxing such selection criteria from $M(j) > 15~\mathrm{GeV}$ to $M(j) > 10~\mathrm{GeV}$ and from $\Delta M < 25~\mathrm{GeV}$ to $\Delta M < 35~\mathrm{GeV}$, we notice from Tab.~\ref{tab:cross_sections_2} that both benchmark scenarios can now be potentially probed at both run-III and HL-LHC. Although, note that for benchmark H6, only one event is expected at run-III.
\begin{table}[htb!]
\centering
\captionsetup{justification=raggedright,singlelinecheck=false}
\resizebox{\columnwidth}{!}{%
	\begin{tabular}{c|c|c|c|c|c}
			& $\sigma$ (before cuts, in fb) & $\sigma$ (after cuts, in fb) & Events at run-III & Events at HL-LHC & NN events\\ \hline
			Signal (Point H1) &	
			\makecell{
				\begin{math}
				0.0594
				\end{math}
			}
			&
			0.0065
			&
			2
			&
			19
            &
            $\sim 19.00$
		    \\
		    \hline
			Signal (H2 to H5) &
			\makecell{
            --
			}
			&
			--
			&
			--
			&
			--
            &
            {--}
		    \\
		    \hline
			Signal (Point H6) &
			\makecell{
				\begin{math}
				0.16
				\end{math}
			}
			&
		    0.000699
			&
			$< 1$
			&
			2
            &
            $\sim 2.00$
			\\
			\hline 
		    $\mathrm{Z^0}+\mathrm{jets}$ &
			\makecell{
				\begin{math}
				4.12\times 10^{6}
				\end{math}
			}
			&
			9.64
			&
			2891
			&
			28915
			\\
			\cline{1-5} 
			$t\bar{t}$  &
			\makecell{
				\begin{math}
				{9.85\times 10^{5}}
				\end{math}
			}
			&
			59.18
			&
			17754
			&
			177540
            &
            \multirow{3}{*}{\makecell{H1 $\sim 0.51$ \\ H6 $\sim 0.56$}}
			\\
			\cline{1-5}      
			Single top  &
			\makecell{
				\begin{math}
				3.43\times 10^{5}
				\end{math}
			}
			&
			34.68
			&
			4306
			&
			43068
			\\
			\cline{1-5}
			$t\bar{t}+V$  &
			\makecell{
				\begin{math}
				33.41
				\end{math}
			}
			&
			0.024
			&
			7
			&
			71
			\\
			\cline{1-5}
			Diboson  &
			\makecell{
				\begin{math}
				7.79\times 10^{4}
				\end{math}
			}
			&
			0.045
			&
			13
			&
			135
			\\
			\hline
	\end{tabular}}
	\captionsetup{justification=raggedright,singlelinecheck=false}
	\caption{The predicted total cross section (in fb) for both the signal and each respective background, before and after the selection cut. In the last three columns we indicate the number of expected events following the imposition of the selection criteria, calculated as $N=\sigma \mathcal{L}$ for run-III and the HL phase of the LHC, whereas for the last column we showcase the total number of events for signal and backgrounds based on the NN cut that provided the maximal significance. For the benchmarks H2 to H5, no events survive after the imposition of the selection criteria, with $M(j) > 15~\mathrm{GeV}$ and $\Delta M < 25~\mathrm{GeV}$.}
	\label{tab:cross_sections}
\end{table}
\begin{table}[htb!]
\centering
\captionsetup{justification=raggedright,singlelinecheck=false}
\resizebox{\columnwidth}{!}{%
	\begin{tabular}{c|c|c|c|c|c}
			& $\sigma$ (before cuts, in fb) & $\sigma$ (after cuts, in fb) & Events at run-III & Events at HL-LHC & NN events \\ \hline
			Signal (Point H1) &	
			\makecell{
				\begin{math}
				0.0594
				\end{math}
			}
			&
			0.028
			&
			8
			&
			87
            &
            $\sim 81.73$
		    \\
		    \hline
			Signal (H2 to H5) &
			\makecell{
            --
			}
			&
			--
			&
			--
			&
			--
            &
            --
		    \\
		    \hline
			Signal (Point H6) &
			\makecell{
				\begin{math}
				0.16
				\end{math}
			}
			&
		    0.0048
			&
			1
			&
			14
            &
            $\sim 14.00$
			\\
			\hline 
		    $\mathrm{Z^0}+\mathrm{jets}$ &
			\makecell{
				\begin{math}
				4.12\times 10^{6}
				\end{math}
			}
			&
			92.25
			&
			27675
			&
			276750
            &
			\\
			\cline{1-5} 
			$t\bar{t}$  &
			\makecell{
				\begin{math}
				9.85\times 10^{5}
				\end{math}
			}
			&
			768.08
			&
			230424
			&
			2304240
            &
            \multirow{3}{*}{\makecell{H1 $\sim 3.24$ \\ H6 $\sim 4.00$}}
			\\
			\cline{1-5}      
			Single top  &
			\makecell{
				\begin{math}
				3.43\times 10^{5}
				\end{math}
			}
			&
			301.70
			&
			37470
			&
			374700
            &
			\\
			\cline{1-5}
			$t\bar{t}+V$  &
			\makecell{
				\begin{math}
				33.41
				\end{math}
			}
			&
			0.25
			&
			75
			&
			750
            &
			\\
			\cline{1-5}
			Diboson  &
			\makecell{
				\begin{math}
				7.79\times 10^{4}
				\end{math}
			}
			&
			13.39
			&
			4017
			&
			40170
            &
			\\
			\hline
	\end{tabular}}
	\captionsetup{justification=raggedright,singlelinecheck=false}
	\caption{The predicted total cross section (in fb) for both the signal and each respective background, before and after the selection cut. In the last three columns we indicate the number of expected events following the imposition of the selection criteria, calculated as $N=\sigma \mathcal{L}$ for run-III and the HL phase of the LHC, whereas for the last column we showcase the total number of events for signal and backgrounds based on the NN cut that provided the maximal significance at $\mathcal{L}=3000~\mathrm{fb^{-1}}$. For the benchmarks H2 to H5, no events survive after the imposition of the selection criteria, with $M(j) > 10~\mathrm{GeV}$ and $\Delta M < 35~\mathrm{GeV}$.}
	\label{tab:cross_sections_2}
\end{table}

Of course, these conclusions are taken from a limited sample of benchmark points, and it is possible that other coupling/mass combinations may lead to different conclusions. At any rate, the results from the considered benchmarks already indicate a non-zero number of expected events, which can enable one to set exclusion limits for the production of these scalars in different regions of the phase space.

\section{Statistical significance with Deep Learning}
\label{sec:numerical_results}

The determination of the statistical significance of the considered benchmark points is performed using evolutionary algorithms based on deep learning (DL) methods. The employed methodology is the same as in previous works by some of the authors \cite{Morais:2021ead,Freitas:2020ttd,Bonilla:2021ize}, and thus in what follows we only present the key features.

As detailed in \cite{Freitas:2020ttd}, the inputs of the DL evolutionary method consist of normalised kinematic and angular distributions, which in the current analysis are shown in Fig.~\ref{fig:best_observables}. Additionally, due to the unbalanced nature of the various classes, i.e.~some having more data entries than others, we use a \texttt{SMOTE} algorithm \cite{Chawla_2002} to balance our dataset and use it as input data for the neural networks (NNs). The latter are constructed using \texttt{TensorFlow} \cite{Abadi:2016kic}. 

Last but not least, the calculation of the statistical significance is performed using as metric the Asimov significance, defined as 
\begin{equation}\label{eq:Asimov_sig}
\mathcal{Z_A} = \Bigg[2\Bigg((s + b)\ln\Bigg(\frac{(s+b)(b+\sigma_b^2)}{b^2 + (s+b)\sigma_b^2}\Bigg) -\frac{b^2}{\sigma_b^2}\ln\Bigg(1+\frac{\sigma_b^2 s}{b(b+\sigma_b)}\Bigg)\Bigg)\Bigg]^{1/2},
\end{equation}
where $s$ is the number of signal events, $b$ is the number of background events and $\sigma_b$ is the variance of the background events, with systematic uncertainties of 1 \% \footnote{Based on the European strategy for particle physics (see \cite{CERN_HLLHC}), both ATLAS and CMS, a guideline of 1-1.5\% systematic was set. Additionally, a recent update \cite{ATLAS:2022hro} improved this estimate to 0.83\% based on the collected data from run-II, making a global 1\% a realistic benchmark scenario for both run-III and the HL phase. We varied by a factor $2$ around 1\% systematics and have not found any significant deviations from the calculated values reported in the text.}. 
\begin{figure}
    \centering
    \captionsetup{justification=raggedright,singlelinecheck=false}
    \includegraphics[width=0.49\textwidth]{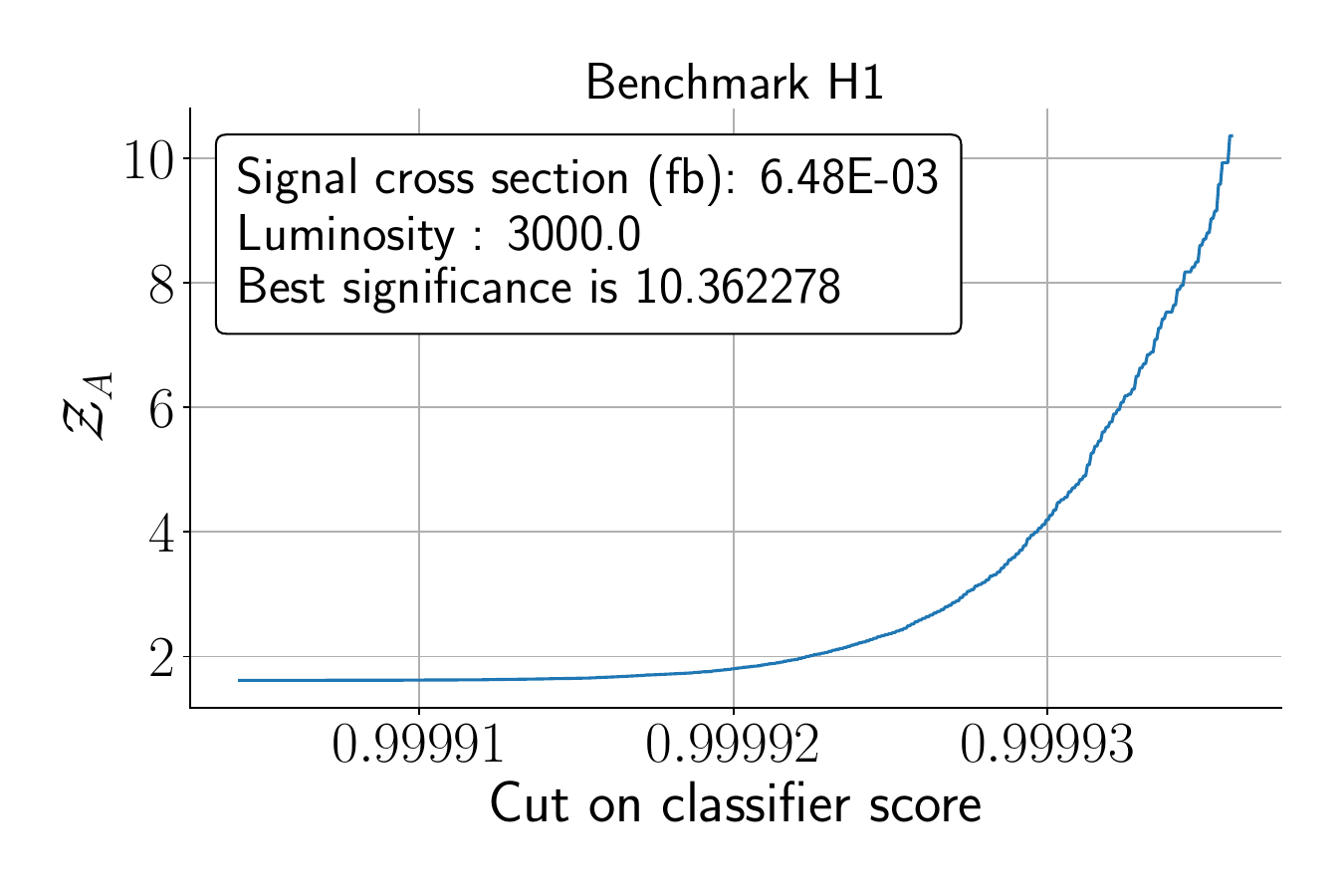} 
    \includegraphics[width=0.49\textwidth]{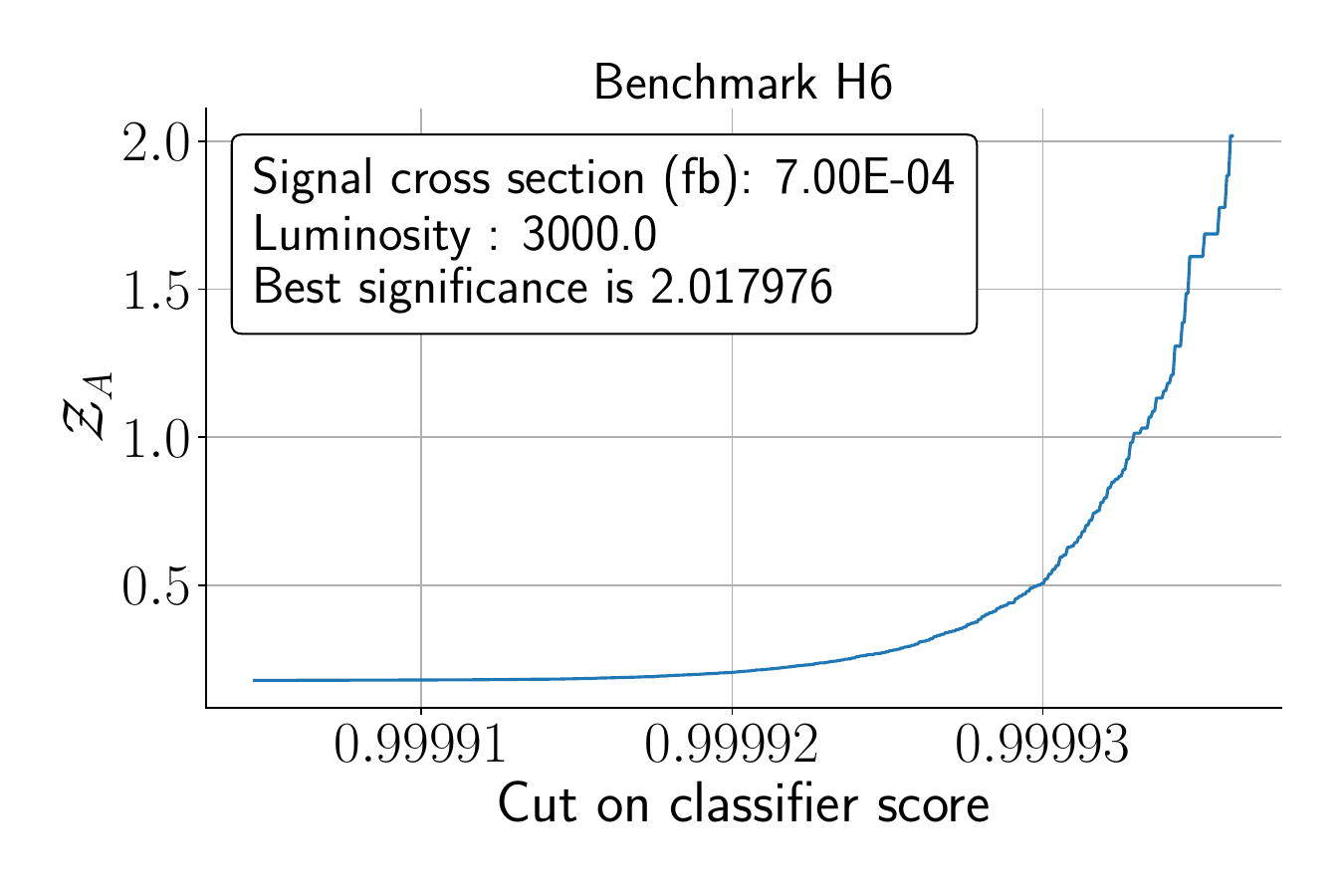}
    \caption{Statistical significance with the mass cuts $M(j) > 15~\mathrm{GeV}$ and $\Delta M < 25~\mathrm{GeV}$. Results are shown for an integrated luminosity of $3000~\mathrm{fb^{-1}}$ and for the NN architecture indicated in appendix~\ref{app:NN_architecture}.}
    \label{fig:Sig_plots_1}
\end{figure}

\begin{figure}
    \centering
    \captionsetup{justification=raggedright,singlelinecheck=false}
    \includegraphics[width=0.49\textwidth]{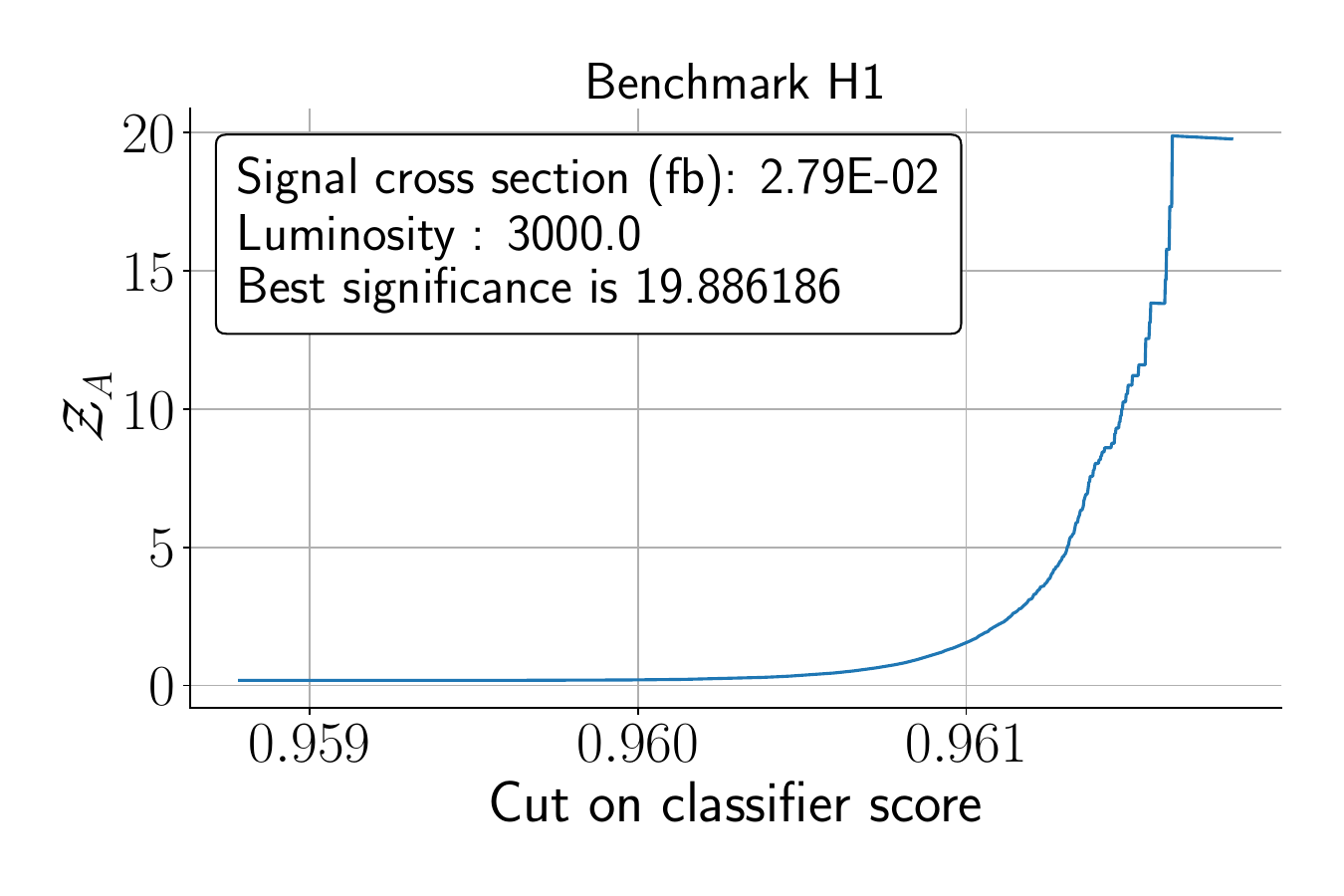} 
    \includegraphics[width=0.49\textwidth]{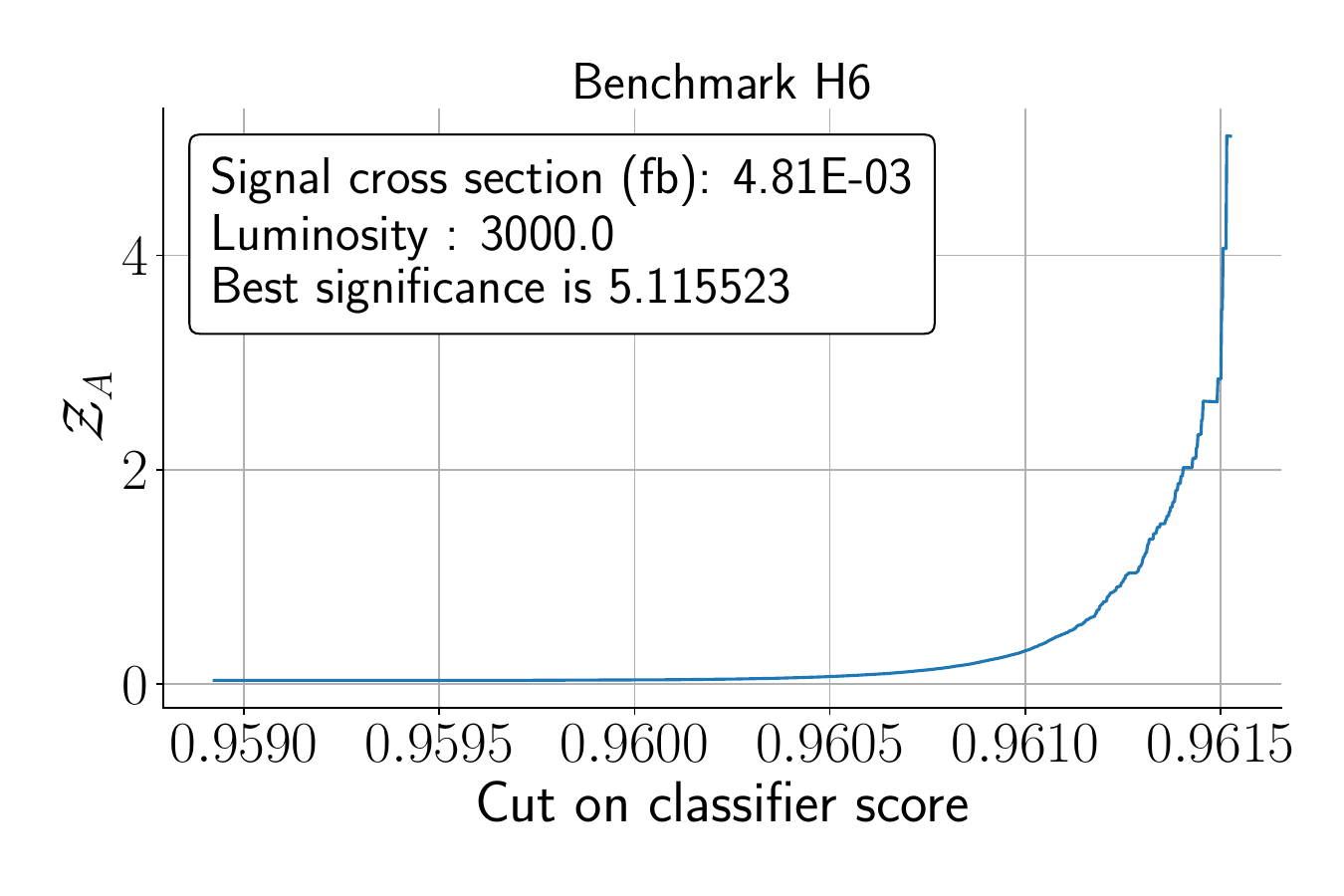} \\
    \caption{Statistical significance with the mass cuts $M(j) > 10~\mathrm{GeV}$ and $\Delta M < 35~\mathrm{GeV}$. Results are shown for an integrated luminosity of $3000~\mathrm{fb^{-1}}$ and for the NN architecture indicated in appendix~\ref{app:NN_architecture}.}
    \label{fig:Sig_plots_2}
\end{figure}

In what follows we determine the statistical significance for our benchmark scenarios, considering first the scenario H1 and the HL phase of the LHC. We show the result in terms of the cut on the classifier score in \cref{fig:Sig_plots_1}. The numerical values for the employed cuts are determined by the NN and can be interpreted as a label for identifying whether a set of features can be classified either as signal or as a background event. From here, we also indicate the best Asimov significance obtained for each of the metrics, whose values for both of the employed cuts read as
\begin{equation}\label{eq:sig}
\begin{aligned}
&\underline{(M(j) > 15~/~\Delta M < 35~\mathrm{GeV})}: \mathcal{Z}_A = 19.89\sigma\,, \\ \\
&\underline{(M(j) > 10~/~\Delta M < 25~\mathrm{GeV})}: \mathcal{Z}_A = 10.36\sigma\,,
\end{aligned}
\end{equation}
Notice that, for both cases, we obtain a statistical significance well beyond $5\sigma$, suggesting that the benchmark H1 is a good candidate to be tested at the LHC. 

One can also confirm that more lenient constraints lead to a greater significance. Indeed, while more restrictive constraints on the phase space result in further reduced backgrounds, the expected number of events for the signal becomes larger when more lenient cuts are considered. Combining this with the good separation obtained between the background and the signal, see \cref{fig:BDT_discr}, the overall significance then becomes larger. For the case of the benchmark H6, at the HL phase, the statistical significance reads as
\begin{equation}\label{eq:sig_1}
\begin{aligned}
&\underline{(M(j) > 15~/~\Delta M < 35~\mathrm{GeV})}: \mathcal{Z}_A = 5.12\sigma\,, \\ \\
&\underline{(M(j) > 10~/~\Delta M < 25~\mathrm{GeV})}: \mathcal{Z}_A = 2.02\sigma\,,
\end{aligned}
\end{equation}
from where we conclude that for the more restrictive cuts, we can only probe this point with a significance of {\color{red} $2.02\sigma$}, whereas for more lenient ones we obtain a statistical significance of {\color{red} $5.12\sigma$}. We have also found that this method works more efficiently for larger masses of the scalar fields inside the decay chain, as is the case of H1 in comparison to H6. This results from the fact that, for the lighter scalars' scenarios, the cut on the invariant masses of the light jets can severely reduce the cross section. 

For completeness, we also extended this analysis into luminosities of $300~\mathrm{fb}^{-1}$ and $1000~\mathrm{fb}^{-1}$ presenting the results in \cref{tab:significance_table}. In particular, we have found that the benchmark H1 can already be tested at run-III, with a significance of up to $5.48\sigma$ or $1.96\sigma$, whether we consider the less restrictive or the more lenient selection criteria, respectively. For benchmark H6, on the other hand, it can only be tested at the HL phase of the LHC.
\begin{table}[htb!]
\begin{center}
\captionsetup{justification=raggedright,singlelinecheck=true}
\begin{tabular}{c|c|l|l|cll|c|l|l|c|l|l}
Benchmarks & \multicolumn{1}{c|}{\makecell{H1 \\ $M(j)>15~\mathrm{GeV}$/$\Delta M < 25~\mathrm{GeV}$}} & \multicolumn{1}{c|}{\makecell{H6 \\ $M(j)>15~\mathrm{GeV}$/$\Delta M < 25~\mathrm{GeV}$}} \\[2mm] \hline
\multirow{2}{*}{$300~\mathrm{fb^{-1}}$ \hspace{0.3em}$\mathcal{Z}_A$} &  \multirow{2}{*}{$1.96\sigma$} & \hspace{5.3em} \multirow{2}{*}{$0.27\sigma$}  \\[5mm] \hhline{=============}
\multirow{2}{*}{$1000~\mathrm{fb^{-1}}$ \hspace{0.3em}$\mathcal{Z}_A$} &  \multirow{2}{*}{$4.81\sigma$} & \hspace{5.3em} \multirow{2}{*}{$0.80\sigma$}  \\[5mm] \hhline{=============}
\multirow{2}{*}{$3000~\mathrm{fb^{-1}}$ \hspace{0.3em}$\mathcal{Z}_A$} &  \multirow{2}{*}{\redBU $10.36\sigma$} & \hspace{5.3em} \multirow{2}{*}{$2.02\sigma$}  \\[5mm] \hhline{=============}
\end{tabular} 
\begin{tabular}{c|c|l|l|cll|c|l|l|c|l|l}
Benchmarks & \multicolumn{1}{c|}{\makecell{H1 \\ $M(j)>10~\mathrm{GeV}$/$\Delta M < 35~\mathrm{GeV}$}} & \multicolumn{1}{c|}{\makecell{H6 \\ $M(j)>10~\mathrm{GeV}$/$\Delta M < 35~\mathrm{GeV}$}} \\[2mm] \hline
\multirow{2}{*}{$300~\mathrm{fb^{-1}}$ \hspace{0.3em}$\mathcal{Z}_A$} &  \multirow{2}{*}{\redBU $5.48\sigma$} & \hspace{5.3em} \multirow{2}{*}{$1.29\sigma$}  \\[5mm] \hhline{=============}
\multirow{2}{*}{$1000~\mathrm{fb^{-1}}$ \hspace{0.3em}$\mathcal{Z}_A$} &  \multirow{2}{*}{\redBU $11.49\sigma$} & \hspace{5.3em} \multirow{2}{*}{$2.95\sigma$}  \\[5mm] \hhline{=============}
\multirow{2}{*}{$3000~\mathrm{fb^{-1}}$ \hspace{0.3em}$\mathcal{Z}_A$} &  \multirow{2}{*}{\redBU $19.89\sigma$} & \hspace{5.3em} \multirow{2}{*}{\redBU $5.12\sigma$}  \\[5mm] \hhline{=============}
\end{tabular}
\caption{The statistical significance for each of the benchmarks/cuts mentioned in the text. Benchmark points that are not shown indicate that the signal cross section is too low for events to be produced at any given luminosity. In {\redBU red} we indicate points that pass the $5\sigma$ threshold for discovery.}
\label{tab:significance_table}
\end{center}
\end{table}

\section{Conclusion}\label{sec:Conclusion}

In this article, we have conducted a collider phenomenological analysis using as an example model a BGL version of the NTHDM. We have focused on a potential signal topology characterized by two charged leptons and four jets in its final state. Based on Monte-Carlo generated datasets, we have demonstrated that one can use the mass-difference information between two pairs of the jets in order to efficiently reconstruct all intermediate particles, while simultaneously damping the main irreducible backgrounds, such as $\mathrm{Z^0+jets}$. We have used the \texttt{TMVA} framework to identify the ten best observables that offer the greatest discriminating power between signal and background distributions. These were subsequently used as features in a Deep-Learning evolution algorithm engineered to find the Neural Network that best optimizes the statistical significance of a hypothetical discovery. We have employed such methods on pre-selected benchmark points, which are consistent with electroweak and flavour observables' constraints, and whose cross sections are high enough for it to be accessible at future collider experiments. In this regard, we found that one can potentially test the considered model with a statistical significance greater than 5 standard deviations, for both the LHC run-III and its future HL upgrade, depending on the benchmark and employed selection criteria. 

More importantly, it is worth mentioning that the main take-away message of this study is that both the methods and the considered signal topology are model-independent and can be applied to any multi-scalar scenario featuring, at least, three new physical neutral Higgs bosons allowed to couple in a triple vertex interaction.
\\ \\
{\bf Acknowledgements.}
We thank Felipe F. Freitas for the fruitful discussions during the initial stages of this project.
J.G., A.P.M. and V.V. are supported by the Center for Research and Development in Mathematics and Applications (CIDMA) through the Portuguese Foundation for Science and Technology (FCT - Funda\c{c}\~{a}o para a Ci\^{e}ncia e a Tecnologia), references UIDB/04106/2020 and UIDP/04106/2020. A.P.M., J.G. and V.V are supported by the projects PTDC/FIS-PAR/31000/2017 and CERN/FIS-PAR/0021/2021. 
A.P.M.~is also supported by national funds (OE), through FCT, I.P., in the scope of the framework contract foreseen in the numbers 4, 5 and 6 of the article 23, of the Decree-Law 57/2016, of August 29, changed by Law 57/2017, of July 19.
J.G is also directly funded by FCT through a doctoral program grant with the reference 2021.04527.BD.
R.P.~is supported in part by the Swedish Research Council grant, contract number 2016-05996, as well as by the European Research Council (ERC) under the European Union's Horizon 2020 research and innovation programme (grant agreement No 668679). 
P.F. is supported by CFTC-UL under FCT contracts UIDB/00618/2020, UIDP/00618/2020, and by
the projects CERN/FIS-PAR/0002/2017 and CERN/FIS-PAR/0014/2019.
A.O. is supported by the FCT project CERN/FIS-PAR/0029/2019.
\appendix

\section{Neural network architectures}\label{app:NN_architecture}

\begin{table*}[htb!]    
	\centering
    \captionsetup{justification=raggedright,singlelinecheck=false}
	\resizebox{1.0\textwidth}{!}{\begin{tabular}{|c|c|c|c|c|}
		\hline
		\midrule
		\makecell{Architecture}  & \makecell{\underline{Neurons} :  256 for input and hidden layers and 6 for output; \\ \underline{Number of layers}:  1 input + 3 hidden + 1 output; \\ \underline{Regularizer}: L1L2; \\
		\underline{Initializer} : For input and hidden layers, VarianceScaling with normal distribution \\ and in fan in mode. Output layer, VarianceScaling with uniform distribution\\  and in fan avg mode;\\
		\underline{Activation functions}: tanh for input/hidden layers. Sigmoid for output layer \\ 
		\underline{Optimizer}: Adam \\		} 
		\\
		\hline
	\end{tabular}}
	\caption{NN architectures that were found by the evolutionary algorithm for the benchmarks H1/H6. for the selection criteria $M(j) > 10~\mathrm{GeV}$ and $\Delta M < 35~\mathrm{GeV}$.}
	\label{tab:Network_EVO_1}
\end{table*}

\begin{table*}[htb!]    
	\centering
    \captionsetup{justification=raggedright,singlelinecheck=false}
	\resizebox{1.0\textwidth}{!}{\begin{tabular}{|c|c|c|c|c|}
		\hline
		\midrule
		\makecell{Architecture}  & \makecell{\underline{Neurons} :  1024 for input and hidden layers and 6 for output; \\ \underline{Number of layers}:  1 input + 2 hidden + 1 output; \\ \underline{Regularizer}: L1L2; \\
		\underline{Initializer} : For input and hidden layers, VarianceScaling with uniform distribution \\ and in fan in mode. Output layer, VarianceScaling with uniform distribution\\  and in fan avg mode;\\
		\underline{Activation functions}: Sigmoid\\ 
		\underline{Optimizer}: Adam \\		} 
		\\
		\hline
	\end{tabular}}
	\caption{NN architectures that were found by the evolutionary algorithm for the benchmarks H1/H6. for the selection criteria $M(j) > 15~\mathrm{GeV}$ and $\Delta M < 25~\mathrm{GeV}$.}
	\label{tab:Network_EVO_2}
\end{table*}

\newpage

\section{Numerical benchmarks}\label{app:Numerical_benchmarks}

Here, we write down the numerical values for the benchmark points that were studied in this work. The values indicated here are not exact, but rounded to 3 significant figures. For exact values, as well as some complementary information, see  \url{https://github.com/Mrazi09/BGL-ML-project}. Quadratic mass terms are in units of $\mathrm{GeV}^2$ ($\mu_1^2$, $\mu_2^2$, $\mu_3^2$, $\mu_b^2$ and $\mu_S^2$) and masses are in units GeV ($M_{H_2}$, $M_{H_3}$, $M_{A_2}$, $M_{A_3}$, $M_{H^\pm}$). The VEV of the singlet field $S$ is in unit of GeV ($v_S$)

\begin{itemize}
    \item \underline{H1}
    \begin{equation}\label{eq:H1_numbers}
    \begin{aligned}
        &\lambda_1 = 2.413\,, \quad \lambda_2 = 0.104\,, \quad \lambda_3 = 1.515\,, \quad \lambda_4 = 4.224\,, \quad \lambda_1^\prime = 0.0257\,, \\ &\lambda_2^\prime = -2.721\,, \quad \lambda_3^\prime = -0.0397\,, \quad \mu_1^2 = 1.973\times 10^{7}\,, \quad \mu_2^2 = 2.799\times 10^{5}\,, \\
        & \mu_3^2 = -3.132\times 10^{6}\,, \quad \mu_S^2 = -8.430\times 10^6\,, \quad \mu_b^2 = -2.837\times 10^5\,, \\
        &a_3 = 0.433\,, \quad v_{S} = 3781.31\,, \quad M_{H_2} =  599.085\,, \quad M_{H_3} = 907.607\,, \\
        &M_{A_2} = 315.991\,, \quad M_{A_3} = 955.091\,, \quad M_{H^\pm} = 566.363
    \end{aligned}
    \end{equation}
    \item \underline{H2}
    \begin{equation}\label{eq:H2_numbers}
    \begin{aligned}
        &\lambda_1 = 4.123\,, \quad \lambda_2 = 0.133\,, \quad \lambda_3 = -1.229\,, \quad \lambda_4 = 1.996\,, \quad \lambda_1^\prime = 0.330\,, \\ &\lambda_2^\prime = 4.099\,, \quad \lambda_3^\prime = -0.248\,, \quad \mu_1^2 = -1.030\times 10^{6}\,, \quad \mu_2^2 = 6.889\times 10^{4}\,, \\
        & \mu_3^2 = -2.392\times 10^{5}\,, \quad \mu_S^2 = -1.339\times 10^5\,, \quad \mu_b^2 = -6.275\times 10^4\,, \\
        &a_3 = 0.716\,, \quad v_{S} = 244.648\,, \quad M_{H_2} =  286.917\,, \quad M_{H_3} = 741.153\,, \\
        &M_{A_2} = 159.196\,, \quad M_{A_3} = 566.752\,, \quad M_{H^\pm} = 411.940
    \end{aligned}
    \end{equation}
    \item \underline{H3}
    \begin{equation}\label{eq:H3_numbers}
    \begin{aligned}
        &\lambda_1 = 1.689\,, \quad \lambda_2 = 0.124\,, \quad \lambda_3 = 1.414\,, \quad \lambda_4 = 2.782\,, \quad \lambda_1^\prime = 0.195\,, \\ &\lambda_2^\prime = -3.492\,, \quad \lambda_3^\prime = -0.224\,, \quad \mu_1^2 = 2.211\times 10^{6}\,, \quad \mu_2^2 = 1.223\times 10^{5}\,, \\
        & \mu_3^2 = -4.958\times 10^{5}\,, \quad \mu_S^2 = -1.216\times 10^5\,, \quad \mu_b^2 = -9.680\times 10^4\,, \\
        &a_3 = 0.808\,, \quad v_{S} = 1065.478\,, \quad M_{H_2} =  527.554\,, \quad M_{H_3} = 724.461\,, \\
        &M_{A_2} = 205.411\,, \quad M_{A_3} = 707.602\,, \quad M_{H^\pm} = 524.699
    \end{aligned}
    \end{equation}
    \item \underline{H4}
    \begin{equation}\label{eq:H4_numbers}
    \begin{aligned}
        &\lambda_1 = 0.914\,, \quad \lambda_2 = 0.129\,, \quad \lambda_3 = -0.0672\,, \quad \lambda_4 = 2.251\,, \quad \lambda_1^\prime = 0.137\,, \\ &\lambda_2^\prime = 0.0990\,, \quad \lambda_3^\prime = -0.166\,, \quad \mu_1^2 = 1.309\times 10^{5}\,, \quad \mu_2^2 = 1.107\times 10^{5}\,, \\
        & \mu_3^2 = -4.037\times 10^{5}\,, \quad \mu_S^2 = -1.181\times 10^5\,, \quad \mu_b^2 = -7.311\times 10^4\,, \\
        &a_3 = 0.536\,, \quad v_{S} = 1184.279\,, \quad M_{H_2} =  397.452\,, \quad M_{H_3} = 705.621\,, \\
        &M_{A_2} = 188.008\,, \quad M_{A_3} = 611.067\,, \quad M_{H^\pm} = 448.519 
    \end{aligned}
    \end{equation}
    \item \underline{H5}
    \begin{equation}\label{eq:H5_numbers}
    \begin{aligned}
        &\lambda_1 = 2.699\,, \quad \lambda_2 = 0.138\,, \quad \lambda_3 = -0.497\,, \quad \lambda_4 = 0.756\,, \quad \lambda_1^\prime = 0.0401\,, \\ &\lambda_2^\prime = -0.442\,, \quad \lambda_3^\prime = -0.0673\,, \quad \mu_1^2 = 1.096\times 10^{6}\,, \quad \mu_2^2 = 1.540\times 10^{5}\,, \\
        & \mu_3^2 = -3.714\times 10^{5}\,, \quad \mu_S^2 = 3.554\times 10^4\,, \quad \mu_b^2 = -2.270\times 10^5\,, \\
        &a_3 = 0.152 \,, \quad v_{S} = 2190.371\,, \quad M_{H_2} =  215.563\,, \quad M_{H_3} = 626.921\,, \\
        &M_{A_2} = 177.492\,, \quad M_{A_3} = 683.604\,, \quad M_{H^\pm} = 156.890 
    \end{aligned}
    \end{equation}    
    \item \underline{H6}
    \begin{equation}\label{eq:H6_numbers}
    \begin{aligned}
        &\lambda_1 = -3.480\,, \quad \lambda_2 = 0.110\,, \quad \lambda_3 =  -0.489\,, \quad \lambda_4 = 2.766\,, \quad \lambda_1^\prime = 0.00574\,, \\ &\lambda_2^\prime = -1.839\,, \quad \lambda_3^\prime = -0.0510\,, \quad \mu_1^2 = 1.336\times 10^{7}\,, \quad \mu_2^2 = 3.634\times 10^{5}\,, \\
        & \mu_3^2 = -1.899\times 10^{6}\,, \quad \mu_S^2 = 1.454\times 10^5\,, \quad \mu_b^2 = -2.275\times 10^5\,, \\
        &a_3 = 0.260 \,, \quad v_{S} = 3794.448 \,, \quad M_{H_2} =  402.807\,, \quad M_{H_3} = 429.589\,, \\
        &M_{A_2} = 217.359\,, \quad M_{A_3} = 772.323\,, \quad M_{H^\pm} = 330.882 
    \end{aligned}
    \end{equation}

\end{itemize}

\bibliography{bib}
\end{document}